\documentclass[preprint]{aastex}
 
\usepackage{verbatim}
\usepackage{amsmath}
 
\def\persqcm{$\rm cm^{-2}$}

\def\h2{$\rm H_2$}
\def\error{$\pm$}
\def\e#1{$\times 10^{#1}$}
\def\tenup#1{10$^{#1}$}
\def\asec{\arcsec}

\def\deg{\arcdeg}

\def\kms{km~s$^{-1}$}

\def\solmass{$\rm M_{\sun}$}

\newcommand{\convunits}{$\rm cm^{-2}\,(K\,km\,s^{-1})^{-1}$}

\newcommand{\jybkms}{$\rm Jy\,b^{-1}\,km\,s^{-1}$}

\newcommand{\mjb}{$\rm mJy\:beam^{-1}$}
\newcommand{\mjysr}{$\rm MJy\:sr^{-1}$}

\newcommand{\msunsqpc}{\solmass~pc$^{-2}$}
\newcommand{\bit}{\begin{itemize}}
\newcommand{\eit}{\end{itemize}}
\newcommand{\beamsz}[2]{${#1}'' \times {#2}''$}
\newcommand{\mum}{\micron}
\newcommand{\Mjysr}{MJy~sr$^{-1}$}
                                                                                
\begin{document}
                                                                                
\title{Mid- to Far-IR Emission and Star Formation in Early-Type Galaxies}
\author{Lisa M.\ Young}
\affil{Physics Department, New Mexico Tech, Socorro, NM 87801}
\email{lyoung@physics.nmt.edu}
\author{George J.\ Bendo}
\affil{Astrophysics Group, Imperial College London, Blackett Laboratory, 
   Prince Consort Road, London, W5 2JN, United Kingdom}
\author{Danielle M.\ Lucero}
\affil{Physics Department, New Mexico Tech, Socorro, NM 87801}

\begin{abstract}
Many early-type galaxies have been detected at wavelengths of 24 to 160\mum\ but the
emission is usually dominated by heating from an AGN or from the
evolved stellar population. Here we present {\it Spitzer} MIPS
observations of a sample of elliptical and lenticular galaxies that
are rich in cold molecular gas, and we investigate whether the MIR to
FIR emission could be associated with star formation activity.  The
24\mum\ images show a rich variety of structures, including nuclear
point sources, rings, disks, and smooth extended emission.
Comparisons to matched-resolution CO and radio continuum images
suggest that the bulk of the 24\mum\ emission can be traced to star
formation with some notable exceptions.  The 24\micron\ luminosities
of the CO-rich galaxies are typically a factor of 15 larger than what
would be expected from the dust associated with their evolved stars.
In addition, FIR/radio flux density ratios are consistent with star
formation. 
We conclude that the star formation rates in
$z=0$ elliptical and lenticular galaxies, as inferred by other authors from UV and
optical data, are roughly consistent with the molecular gas abundances
and that the molecular gas is usually unstable to star formation
activity.
\end{abstract}


\section{Introduction}

In recent years, UV and optical photometry and spectroscopy of nearby
elliptical galaxies has suggested that these galaxies, which have a
reputation for being old, red, and dead, may not be quite as dead as
previously assumed.  Between a few percent to 30\% of local
ellipticals appear to be experiencing low levels of ongoing star
formation activity
\citep{schawinski07a,schawinski07b,kaviraj07,fukugita04,rogers07}.
The star formation is not intense enough to affect
the galaxies' morphological classification, as it only amounts to a
few percent of the total stellar mass.  However, this disk growth
inside spheroidal galaxies may be a faint remnant of a process that
was more vigorous in the past and may have played a role in
establishing the range of galaxy morphologies we observe today.

Star formation, of course, requires cold gas, so interpreting the UV
and optical data in terms of star formation activity has important
implications both for the early-type galaxies and for a general
understanding of the star formation process.  It is not obvious that
star formation should ``work'' the same way inside spheroidal galaxies
as it does inside disks, with the same efficiency, the same dependence
on the gas surface density, or the same regulatory mechanisms.
For example, it has been hypothesized that even if there is a
molecular disk inside an elliptical or lenticular galaxy, the disk
would probably be stabilized by the galaxy's steep gravitational
potential \citep{kennicutt89,okuda,kawata07}.  Thus it is of interest
to probe the relationships between molecular gas and star formation
activity in early-type galaxies.

It is not as straightforward, however, to measure star formation rates
in early-type galaxies as it is in spirals.  Nebular line emission is
common in ellipticals \citep{shields91,goudfrooij94,sarzi06}, but
it is usually not thought to be associated with star formation.  Its 
distribution is generally smooth, centrally peaked, and sometimes filamentary
\citep{shields91,macchetto96}; its morphology and line ratios suggest ionization
sources such as AGN activity, evolved stars, or cooling from the hot gas
phase rather than star formation.
Far-IR (FIR) and cm-wave
radio continuum emission are also commonly used as tracers of star
formation activity in gas-rich spirals, but AGNs and the evolved stellar
population have been identified as the sources of mid-IR (MIR) and FIR
emission in most elliptical galaxies \citep{temi07,temi08}.

Here we investigate evidence for and against star formation activity
in a sample of elliptical and lenticular galaxies that have unusually
large molecular gas contents.  We make use of matched-resolution
images of the molecular gas distribution, the cm-wave radio continuum
and the 24\mum\ intensity.  The properties of more typical, CO-poor
early-type galaxies are reviewed so that they can provide a comparison
sample for the CO-rich early-type galaxies.  We then present the
morphology of the 24\micron\ emission, with simple parametric fits and
comparisons to molecular gas, radio continuum, and optical images that
show dust disks in silhouette.  The mid-IR and far-IR flux densities
of the CO-rich galaxies are presented along with discussion of the
24\micron\ to 2\micron\ flux density ratios and the FIR/radio flux
density ratios as diagnostics of whether the mid- to far-IR emission
has a circumstellar or star formation origin.  The results are mixed;
in some cases it is clear that the 24\micron\ emission is primarily
associated with star formation, and in other cases heating by the evolved
stellar population or even by an AGN are inferred.  Thus, the results
are at least qualitatively consistent with the suggestions that
present-day star formation activity may be occurring in substantial
numbers of local early-type galaxies.

\section{Context: the Mid- and Far-Infrared Continuum Emission of 
Early-Type Galaxies}

The ISOCAM and ISOPHOT instruments aboard the {\it Infrared Space
Observatory} provided some insight into the variety of processes that
contribute to the mid-IR and far-IR emission of early-type galaxies.
For example, a handful of elliptical and lenticular galaxies found their way
into the {\it ISO} Atlas of Bright Spiral Galaxies and are discussed
by \citet{bendo02a,bendo02b} with particular reference to the rate and
distribution of star formation activity.  The 12\micron\ images show very
little emission beyond their galaxies' nuclei.  In \citet{bendo02b}, enhanced MIR and FIR to $K$
flux density ratios are used as star formation indicators.  It is not
at all obvious that this interpretation is accurate for early-type
galaxies, but in most of these cases the ratios are consistent with
interstellar radiation fields due to the old stellar population so
little if any star formation activity is inferred.  There are a
minority of E-S0/a galaxies whose FIR/$K$ flux density ratios (within
a 15\asec\ aperture) are as high as the median values for Sb-Scd
spirals, suggesting the possibility of star formation activity.

\citet{xilouris04} fit the spectral energy distributions
(SEDs) and compared 15\micron\ images to optical and near-IR images
for a sample of 18 ellipticals, dwarf ellipticals, and lenticulars.
In two cases they found no evidence for excess 15\micron\ emission
over the stellar photospheric emission; in most of the rest of the
sample there is such an excess and it is smoothly distributed, more or
less following the stellar distribution.  A few targets show thermal
emission at 15\micron\ from the dust that is visible in silhouette in
the optical images, and two show AGN emission (a nuclear point source
and even some synchrotron radiation from the jet in M87).  Thus, in
early-type galaxies the stellar photospheres, circumstellar dust,
silhouette dust lanes or disks, AGN, and (possibly) star formation
all contribute in the mid-IR images.

Fourteen galaxies classified as E, S0, or S0/a were
observed as part of the Spitzer Infrared Nearby Galaxies Survey
\citep[SINGS;][]{ketal03}.  Eleven of these were detected up to
160\mum\, and the 24\micron\ morphologies of these galaxies are
discussed by Bendo et al.\ (2007).  With a couple of exceptions, such
as NGC~1316 (which has asymmetric extended emission over 2\arcmin) and
NGC~5866 (which contains an edge-on disk), these early-type galaxies
tended to have only poorly resolved nuclear emission 
at 24\mum.  Six of the 11 also have published CO
observations \citep{wch,lees91,knapp96,ws03}, and four of the six are
detected. NGC~5866 is notably CO-rich, as a large amount of molecular
gas (4.4\e{8} \solmass) is very clearly detected by \citet{ws03}.
However, maps of the molecular gas distributions in most of these
galaxies are not currently available.

MIPS observations of more typically CO-poor elliptical galaxies have
also been published by \citet{kaneda} and \citet{temi07,temi08}.  Of
the 19 galaxies analyzed by \citet{temi08}, 13 have been searched for
CO emission and none have been detected \citep{combes07, swy07}.
\citet{temi08} show that the 24\mum\ emission from their ellipticals
follows the $r^{1/4}$ near-IR surface brightness profiles very closely. 
The 24\micron\ emission even has the same effective radius as in $K$;
the ratio of two radii is found to be 0.96 with a
dispersion of 0.20.  In addition, \citet{temi07} have shown that the
24\mum\ emission globally tracks the optical luminosity in elliptical
galaxies as there is a tight linear correlation between the 24\mum\
flux density and the $B$-band flux density.  This 24\mum\ emission is
interpreted to be circumstellar dust from the mass loss of post main
sequence stars.

In short, several processes may contribute to MIR emission from
early-type galaxies.  In the majority of the elliptical galaxies that
are not CO-rich the 24\mum\ emission seems to either follow the
stellar photospheric emission or a nuclear source.  Extended
15\micron\ and 24\micron\ emission has been observed from a few
early-type galaxies that are very rich in molecular gas or that
exhibit silhouetted dust.  Until now, however, it has been rare to be
able to compare the distribution of possible star formation activity
in early-type galaxies to that of the molecular gas, the raw material
for the star formation.

\section{Target Selection, Observations, and Data Reduction}\label{reduction}

Most early surveys for molecular gas in early-type galaxies were
strongly biased towards FIR-bright targets, with a typical selection
criterion having an IRAS 100\micron\ flux density $>$ 1 Jy
\citep{lees91,wch,knapp96}.  More recent CO searches are not
FIR-biased, but they still find significant molecular gas contents.
\citet{ws03} reached a surprisingly high CO detection rate of $78\%$
in a volume-limited sample of nearby field lenticular galaxies, and
\citet{swy07} detected CO emission in $33\%$ of a similar sample of
field ellipticals.  \citet{combes07} also detected CO emission in
$28\%$ of the early-type galaxies in the SAURON survey
\citep{dezeeuw02}, a representative sample that uniformly fills an
optical magnitude -- apparent axis ratio space.  Thus, the CO
detection rates in ellipticals and lenticulars are high enough to
support the UV-inferred incidence of star formation activity (if that
gas does indeed engage in star formation).  The cold gas masses are
highly variable in these detections, with $M_{\rm gas}/L_B$ in the
range $10^{-1}$ to $10^{-3}$ and lower.

Since we are interested in morphology as a means of distinguishing the
origin of the MIR, FIR, and radio emission, we have selected for this
project some of the relatively few elliptical and lenticular galaxies
with maps resolving their molecular gas distribution
\citep{young02,young05,ybc08}.  Because of the way galaxies were
selected for CO mapping, the targets were already known to be
FIR-bright and to have concentrations of molecular gas in their
centers (as opposed to their outskirts).  We also apply a criterion on
the angular extent of the molecular gas to be able to test the
correspondence between the molecular gas and 24\micron\ emission.  If
the 24\mum\ emission arises in star formation activity, we expect it
to trace the molecular gas.  If the 24\mum\ emission is related to AGN
activity, we expect it to be a point source, and if it comes from
circumstellar dust, it should trace the stellar distribution.  Thus,
we required the targets to have molecular gas in structures on the
order of 20\asec\ to 30\asec\ or more in diameter.  Corresponding dust
emission, if present, should be resolved in the 24\mum\ images.  The
targets' distances range up to 80 Mpc and optical luminosities are in the
range $-21.7 \leq M_B \leq -18.3$ (Table \ref{sampletable}).

Observations of UGC~1503, NGC~807, NGC~2320, NGC~3032, NGC~3656,
NGC~4476, and NGC~5666 were made with the MIPS instrument at 24, 70,
and 160\micron\ in project 20780 of cycle GO-2.  The data were taken
in photometry mode using the large field size in all cases.
Relatively short exposures were used to avoid saturation on these
bright sources in medium-high background regions.  Exposures were made
in four cycles of 10 seconds at 24\micron, with sky offsets of
300\asec, and in 8 cycles of 3 seconds at 160\micron.  Observations at
70\micron\ were made in the fixed cluster-offsets mode with offsets
$\pm$ 80\asec, and 8 cycles of 3 seconds.  To maximize the
morphological information recoverable at 70\asec\ the fine pixel scale
was used.  However, most sources were not expected to be usefully
resolved at 160\micron\, so the single pointing (rather than a raster
map) was used at the longer wavelength.

Data for NGC~4526 were obtained from GTO project 69 (PI: Fazio).
Those observations consist of a 10s exposure at 24\micron\ using the
small field with a 300\asec\ sky offset, 3 cycles of 3 seconds at
70\micron\ using the default pixel scale and small field, and 3 cycles
of 3 seconds at 160\micron\ in the default pixel scale in a 3-by-1
raster map.  Data for NGC~4459 were obtained from GO project 3649 (PI:
C\^ote); the observing modes are similar to those for NGC~4526 except
that the exposure times are 10 cycles of 4 seconds at 24\micron, 10
cycles of 5 seconds at 70\micron, and 10 cycles of 6 seconds at
160\micron.

The 24, 70, and 160~$\mu$m images were created from raw data frames
using the MIPS Data Analysis Tools \citep[MIPS DAT;][]{getal05}
version 3.10 along with additional processing steps.  The processing
steps for the 70 and 160~$\mu$m data are similar, but the steps for
the 24~$\mu$m data are significantly different from these other two
bands.  Therefore, the 24~$\mu$m data processing is described first
followed by descriptions of the 70 and 160~$\mu$m data processing.

The individual 24~$\mu$m frames were first processed through a droop
correction (removing an excess signal in each pixel that is
proportional to the signal in the entire array) and were corrected for
non-linearity in the ramps.  The dark current was then subtracted.
Next, scan-mirror-position dependent flats were created from the data
in each Astronomical Observation Request (AOR) and were applied to the data.  Detector pixels that had
measured signals of 2500 DN s$^{-1}$ in any frame were masked out in the
following three frames so as to avoid having latent images appear in
the data.  Next, a scan-mirror-position independent flat was created
from the data in each AOR and were applied to the data.  Following
this, planes were fit to the zodiacal light emission in the background
regions in each frame (regions falling outside the optical disks of
the galaxies that also did not contain any other bright sources), and
these planes were subtracted from the data.  Next, a robust
statistical analysis was applied in which the values of cospatial
pixels from different frames were compared to each other and
statistical outliers (e.g. probable cosmic rays) are masked out.
After this, a final mosaic was made with pixel sizes of
$1.5^{\prime\prime}$, any residual background in the image was
subtracted, and the data were calibrated into astronomical units.  The
calibration factor for the 24~$\mu$m data is given by \citet{eetal07}
as $(4.54\pm0.18) \times 10^{-2}$ MJy sr$^{-1}$ [MIPS instrumental
unit]$^{-1}$.

In the 70 and 160~$\mu$m data processing, the first step was to fit
ramps to the reads to derive slopes.  In this step, readout jumps and
cosmic ray hits were also removed, and an electronic nonlinearity
correction was applied.  Next, the stim flash frames taken by the
instrument were used as responsivity corrections.  The dark current
was subtracted from the data, and an illumination correction was
applied.  Short term variations in the the signal (often referred to
as drift) were removed from the 70~$\mu$m data; this also subtracted
the background from the data.  Next, a robust statistical analysis was
applied to cospatial pixels from different frames in which statistical
outliers (e.g. pixels affected by cosmic rays) were masked out.  Once
this was done, final mosaics were made using square pixels of
4.5\asec\ for the 70~$\mu$m data and 9\asec\ for the 160~$\mu$m data.
The backgrounds in the 160~$\mu$m data and the residual backgrounds in
the 70~$\mu$m data were measured in regions outside the optical disks
of the galaxies and subtracted, and then flux calibration factors were
applied to the data.  The 70~$\mu$m calibration factors given by
\citet{getal07} are $702 \pm 35$ MJy sr$^{-1}$ [MIPS instrumental
unit]$^{-1}$ for coarse-scale imaging and $2894 \pm 294$ MJy sr$^{-1}$
[MIPS instrumental unit]$^{-1}$ for fine-scale imaging.  The
160~$\mu$m calibration factor is given by \citet{setal07} as
$41.7\pm5$ MJy sr$^{-1}$ [MIPS instrumental unit]$^{-1}$.  An
additional 70~$\mu$m nonlinearity correction given as
\begin{equation} 
f_{70\mu m}(true)=0.581(f_{70\mu m}(measured))^{1.13}
\end{equation} 
by \citet{dggetal07} was applied to coarse-scale imaging data where
the surface brightness exceeded 66 MJy sr$^{-1}$.

\section{Point Spread Functions and Image Fitting}\label{fitting}

For analyzing the radial profiles of the 70 and 160~$\mu$m data and
for creating accurate models of the spatial distribution of 24~$\mu$m
emission, we needed point spread functions (PSFs) that accurately
represent the observed PSFs.  While the STinyTim model \citep{k02} can
be used to model the PSF, the model output does not match the
observed PSF of point sources.  This is because the model assumes that
the PSF is infinitesimally subsampled, whereas the observed PSF is
measured with pixels that effectively blur some of the fine features.
One approach to circumvent this problem is to simply smooth the model
PSF \citep{getal07,setal07}.  As an alternative approach, we created a
set of empirical PSFs from archival data.

The 24 and 70~$\mu$m PSFs were constructed using archival MIPS
photometry mode data from program 20496 (PI: Marscher) of 3C 273, 3C
279, and BL Lac, which are relatively point-like infrared sources at
these wavelengths.  Data from a total of 25 AORs were used.  These
data were processed using the same techniques described in
Section~\ref{reduction} except that the final mosaics were made for
each AOR, and the images axes were left in the native instrumental
rotation instead of being aligned with the J2000 coordinate system.
The final PSFs were created by normalizing the flux densities of every
mosaic to 1 and then median combining the frames from all AORs, which
filtered out extended emission such as a tail-like feature extending
north of 3C 273.

MIPS 160\micron\ photometry mode observations of compact sources
typically do not completely sample the PSF, so we needed to use scan
map data to create a PSF at this wavelength.  In this case, we used
SINGS data of Mrk 33, NGC 1266, NGC 1377, NGC 3265, and NGC 3773 to
create the PSF.  These galaxies were selected because they are
galaxies from SINGS that are unresolved or marginally resolved at
24~$\mu$m, so we can safely assume that the 160~$\mu$m counterparts
are also unresolved.  These data were processed in the same way as the
160~$\mu$m photometry data for the elliptical galaxies in this paper
except that an additional drift removal step was applied to the data,
separate final mosaics were made for each AOR, and the images axes
were left in the native instrumental rotation.  As with the 24 and
70~$\mu$m PSFs, the final PSF was created by normalizing the flux
densities of every mosaic to 1 and then median combining the frames
from all AORs.

The 70\mum\ and 160\mum\ emission from our CO-rich targets is very
poorly resolved, so no detailed morphological analysis was done in
these bands.  However, the 24\micron\ emission is resolved, and simple
model fits are made to parametrize the 24 \mum\ morphologies.  We used
the image fitting techniques described by \citet{bendo-sombrero} to
characterize the spatial distribution of the 24~$\mu$m dust emission
and to measure the flux densities of the galaxies.  These techniques
have been improved in several ways compared to the models that
\citet{bendo-sombrero} applied to the Sombrero Galaxy.  First, we are
now using the empirical PSFs described above instead of the STinyTim
theoretical PSFs.  With these PSFs, we can more accurately model the
region near the centers of these galaxies.  Second, we now treat the
central coordinates of each model component as free parameters.  The
models used in this paper include unresolved point sources, inclined
exponential disks, de Vaucouleurs ($R^{1/4}$) profiles, and rings with
exponential profiles as well as combinations of these, all of which
are convolved with the empirical 24~$\mu$m PSF.

\section{Ancillary Data}

In addition to the CO observations cited above, we also compare the
mid-IR morphology to that of the 1.4 GHz radio continuum, the stellar
distributions from optical and NIR images, and the dust seen in
silhouette against the optical continuum.  The 1.4 GHz radio continuum
emission from galaxies originates in both AGN activity and star
formation \citep{condon92}.  Active nuclei should be distinguishable
as nuclear point sources, perhaps with a jet, whereas kpc-scale
extended emission more likely originates from star formation
\citep{wrobel}.  

Radio continuum images (and one nondetection) for
UGC~1503, NGC~807, NGC~3656, NGC~4476, and NGC~5666 have been
published by \citet{lucero}.  The FIRST survey \citep{FIRST} provides
radio continuum images of NGC~2320, NGC~3032, NGC~4459, and NGC~4526.
All of these 1.4 GHz images have resolutions on the order of 5\asec, which
facilitates comparisons to the MIPS 24\micron\ images at 6\asec\ resolution.
Table \ref{radioims}
lists the beam sizes and rms noise levels of the continuum images.  It also
shows that while the FIRST images have typically a factor of 3 to 4 higher
noise levels than our own, they do not have systematically lower signal-to-noise
ratios.  Thus we do not expect significant biases in the detectability of
extended structures, for example, between the FIRST images and our own 1.4 GHz
data.

Broadband optical $g$ and $i$ images of NGC~3032, NGC~4459, and
NGC~4526 were obtained from the Sloan Digitized Sky Survey (SDSS).
$V$ and $R$ images of UGC~1503, NGC~807, NGC~2320, and NGC~3656 were
obtained by L.\ van Zee in November 2002 with the Mini-Mosaic Imager
(MiniMo) on the WIYN 3.5m telescope and were reduced as described by
\citet{young05}.  The seeing in those images ranged from 0.9\asec\ to
1.3\asec.  $V$ and $I$ images of NGC~4476 and NGC~5666 were obtained
with the Kitt Peak 2.1m telescope and T2KA CCD in April 2003.
Exposure times for those images were 1200 s in $V$ and 1800 s in $I$;
the seeing varied from 1.1\asec\ to 1.6\asec\ in those images and they
were reduced in the same manner as for the WIYN MiniMo images.
Finally, WFPC2 images of NGC~3032, NGC~4459, and NGC~4526 in the F606W
or F555W filters were retrieved from the HST archive along with ACS
observations of NGC~4476.  

The 24\micron\ emission from NGC~807 is also compared with the distribution
of HI in the galaxy.  HI emission was mapped with the National Radio
Astronomy Observatory's Very Large Array\footnote{The National Radio
Astronomy Observatory is operated by Associated Universities, Inc., under
cooperative agreement with the National Science Foundation.} in its C
configuration for 6 hours on 2002 Dec 10 and 6 hours on 2002 Dec 16 
in program AY135.  The total
bandwidth was 3.125 MHz centered at 4650 \kms, giving 63 channels of 21.3 \kms.
The absolute flux scale, bandpass calibration, and time
dependent gain corrections were determined
from observations of the nearby source J0137+331.  All data calibration and image formation
were done using standard calibration tasks in the AIPS package; continuum
emission was subtracted directly from the visibility data by making first
order fits to the line-free channels.  The calibrated data were 
Fourier transformed using Briggs' robust weighting scheme with a robustness
parameter of 0.0, which gave a resolution of
\beamsz{13.8}{13.1} and a rms noise level of 0.2 mJy~beam$^{-1}$.  
The dirty images were cleaned down to a residual level
approximately equal to the rms noise fluctuations.  The integrated intensity
image was made by smoothing, clipping, and then summing the cleaned data cube
in a manner similar to that used by \citet{ybc08}.

\section{Results}

\subsection{Flux Densities and Spectral Energy Distributions}

% 24

None of the CO-rich early-type galaxies are pure point sources at
24\micron.  All are resolved, though they are significantly less
extended than the stellar distributions in the NIR and optical.  Total
flux densities at 24\micron\ are thus derived from large aperture
photometry in the sky-subtracted images.  Unrelated point sources were
first cleaned from the vicinity of the target using the task {\it
imedit} in IRAF, and the total flux densities were summed in a series
of circular apertures up to 60 pixels (90\asec) in radius.  The total
flux densities converge to within $\sim 2\%$ for radii $\geq$ 75\asec,
and therefore Table \ref{fluxes} quotes the flux density within the
75\asec\ aperture.  A color correction is derived based on the power
law index $\beta$ ($S \propto \nu^\beta)$ between the 24\micron\ and
70\micron\ flux densities; the power law indices range from $-1.9$ to
$-3.0$.  These indices are consistent with the mid-IR spectra of
nearby galaxies presented by \citet{smith07}.  The color corrections
tabulated by \citet{setal07} are 0.960 to 0.967.  The dominant
uncertainty in the flux densities is that of the absolute calibration,
roughly 4\% \citep{eetal07}.

% 70 

The 70\micron\ and 160\micron\ emission generally appear unresolved
(Section \ref{70-160}), so we treated these as point sources when
measuring flux densities at these wavelengths.  To select aperture and
color corrections for the 70 and 160~$\mu$m data, we first fit the
60-160~$\mu$m IRAS and {\it Spitzer} data with unmodified blackbody
functions to roughly characterize the SEDs.  The typical color
temperatures derived from these fits were $\sim30$-50~K.  

We choose to measure flux densities in the 70~\mum\ data within an
aperture of 81\asec\ radius, as it is the maximum aperture that stays
entirely within the field-of-view of our 70\micron\ images.
\citet{getal07} do not specifically quote an aperture correction for
these parameters, so we derived that correction factor.  The STinyTim
model \citep{k02} was used to create 60\arcmin -wide model PSFs for
blackbodies with temperatures of 30, 40, and 50~K.  After estimating
the ``background'' level in an annulus of radii 81\asec\ to 100\asec,
we measured the fraction of the total flux density in a circle of
radius 81\asec.  This procedure gives an aperture correction of $1.14
\pm 0.02$.  We also verified that we were able to reproduce the
aperture corrections quoted in \citet{getal07} to within 1\%.  We take
the color correction for a black body of temperature 40 \error\ 10 K
as 0.886 \error\ 0.015 \citep{setal07}.

In most of our targets, the dominant contribution to the uncertainty
in the 70\micron\ flux density is that of the absolute calibration
scale \citep[10\% for fine scale data and 5\% for coarse scale
data,][]{setal07}.  However, the images of NGC~2320 and NGC~4476 still
suffer from some negative artifacts just outside the first Airy ring.
Initial exploration suggests that the 70\micron\ flux densities of
these galaxies may be 20\% and 10\% low, respectively.

% 160 

Flux densities in the 160\micron\ images were measured in an aperture
of radius 64\asec\ with sky background determined in an annulus of
radii 80\asec\ to 160\asec.  For this aperture and a PSF approximately
represented by a 40 K blackbody, we take an aperture correction from
\citet{setal07} as an interpolation between a 30 K and a 50 K
blackbody, which gives 1.3575 \error\ 0.0035.  The color correction,
assuming a blackbody of 40 \error\ 10 K, is 0.964\error 0.010
\citep{setal07}.
For NGC~4459 and NGC~4526, the small fields of view made it difficult
to estimate the background level and flux densities for these two
should be treated with greater caution.  Aside from the background
level and the nonlinearity effect discussed below, the dominant
contribution to uncertainty in the 160\micron\ flux densities is the
absolute calibration scale \citep[12\%,][]{setal07}.

Additional uncertainties in the flux densities come from the fact that
the targets are bright at 70\micron\ and 160\micron.  \citet{getal07}
have shown that aperture photometry gives systematically lower flux
densities than a PSF fitting technique, for sources brighter than
about 1 Jy at 70\micron.  The effect is roughly 5\% at 2 Jy.  The
nonlinearity correction of \citet{dggetal07} (Section \ref{reduction})
should mitigate this effect for the coarse scale data of NGC~4526 and
NGC~4459, but the corresponding correction is not known for fine-scale
data and the effect could be important for NGC~3032, NGC~3656, and
NGC~5666.  \citet{setal07} also show that sources brighter than 2 Jy
at 160\micron\ are underestimated, with the effect being as large as
20\% at 4 Jy.  Finally, we note that since the 160\micron\ emission
for NGC~807 is slightly resolved its aperture correction may be low.

NGC~807 and NGC~5666 also have FIR flux density measurements from the
ISO satellite \citep{temi04}.  In the case of NGC~807, the 160\micron\
flux density of 2.39\error 0.29 Jy that we derive from the MIPS data
is consistent with the 2.8 \error 0.8 Jy measured by Temi et al.\ at
150\micron.  For NGC~5666, our MIPS 160\micron\ flux density
(2.48\error 0.30 Jy) seems low in comparison with the ISO data (3.9 \error\
1.2 Jy at 150\micron\ and 2.7\error 0.8 Jy at 200\micron), but given the
sizes of the uncertainties we cannot conclude that those three values are
inconsistent with each other.
We note that the analysis and processing of the MIPS
160\micron\ data is routine, whereas the ISO data for NGC~807 and
NGC~5666 were obtained with the PHT 32 mode, which is known to suffer
from strong transient effects that make it difficult to calibrate \citep{temi04}.

\begin{figure}
\includegraphics{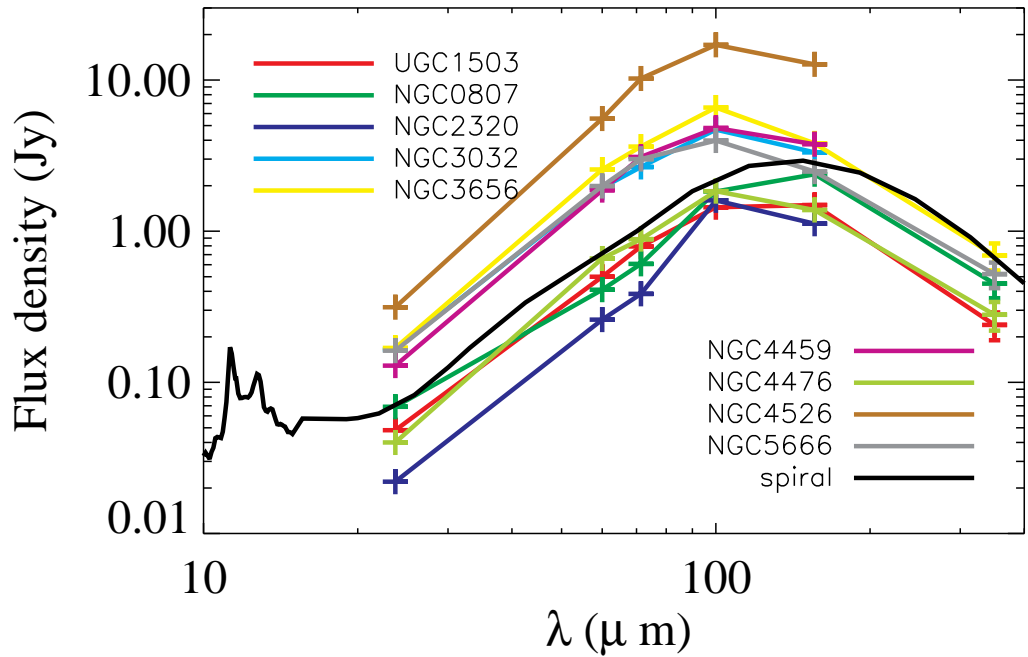}
\caption{Mid- and far-IR flux densities for the CO-rich early-type galaxy sample.
The 350\micron\ points are plotted with appropriate error bars; 
at wavelengths shorter than 350\micron, uncertainties are estimated at 5\% to
20\% (see text) so are comparable to or smaller than the symbol size.  The SED of 
a ``typical spiral" galaxy, from \citet{dggetal07}, is overlaid.
\label{SEDplot}
}
\end{figure}

Far-IR spectral energy distributions for our targets are shown in
Figure \ref{SEDplot}, where we have included the 60\micron\ and
100\micron\ IRAS flux densities and 350\micron\ observations of
\citet{leeuw08} (see Table \ref{q}).  The SEDs are similar to those of
the ``typical'' spiral galaxy from the SINGS survey \citep{dale05,
dggetal07}, but with 70/160\micron\ colors slightly warmer than that
of a typical spiral and 24/70\micron\ colors slightly cooler than the
spiral.  The IRAS 60\micron\ and 100\micron\ flux densities, MIPS
70\micron\ and 160\micron\ flux densities and (where available)
350\micron\ flux densities have been fitted with a blackbody modified
with an emissivity $\propto~\lambda^{-2}$ to characterize the dust
temperatures in these galaxies.  These color temperatures are given in
Table \ref{fluxes}.  We find that the modified blackbody temperatures
of 25-28~K for most of these galaxies are a few degrees warmer than
the average color temperatures for spiral galaxies found by
\citet{de01}, \citet{ptvpm02}, and \citet{bendo03}, which range from
$\sim18$~K to $\sim25$~K depending on the number of thermal components
and combinations of wave bands used.  However, the color temperatures
of the early-type galaxies in this paper are similar to the color
temperatures measured in early-type galaxies by \citet{bendo03} and
\citet{leeuw08}.  In contrast, the color temperatures of 22-23~K
measured for NGC~807 and NGC~2320 are closer to those of spiral
galaxies than to other elliptical and SO galaxies.  

If the dust opacities in the diffuse ISM of most early-type galaxies are low, the dust 
producing the FIR cirrus emission could be bathed in a somewhat more
intense or harder radiation field than is typical for spirals.  This scenario could
explain the higher color temperatures measured in the 60\micron--350\micron\ range.
However, since 24\micron\ emisson primarily traces hot dust associated with
star-forming regions in spiral galaxies, the 24\micron\ emission in these
early-type galaxies could be relatively faint compared to the cirrus emission if
these galaxies contain relatively little star formation compared to their dust
content.

\subsection{24\micron\ Morphology}

In all of the targets, the 24\micron\ emission is resolved.  A variety
of structures (point sources, rings, disks, and $r^{1/4}$ profiles)
are evident in the images and are discussed in greater detail below.
The interpretation of these structures is complicated by the fact that
the 24\mum\ intensity is a function of both the dust surface density
and the illuminating radiation field.  Possible dust heating sources
include the post-main sequence stellar population, star formation
regions, and AGN.  (None of our targets have strong enough radio jets
for spatially resolved synchrotron emission to be important at
24\mum.)  Thus, the 24\mum\ emission by itself may not necessarily
indicate the presence of star formation.  However, evidence from the
CO and radio continuum helps to resolve some of these ambiguities.
For example, star formation is expected to be accompanied by cm-wave
radio continuum emission \citep{condon92}, whereas dust heated by the
radiation from evolved stars would not be.  Therefore the comparisons
with the distribution of the molecular gas (the raw material for star
formation) and the radio continuum provide constraints on the origin
of the 24\mum\ emission.

For each galaxy we consider in detail the evidence for star formation activity,
with a particular view to distinguishing how much
of the 24\micron\ emission is due to star formation.  We take the molecular
gas to be the raw material, so that star formation should only be found within the
CO disks.  We ask whether the radio continuum emission is resolved on scales
similar to the molecular gas, which (if true) suggests that the radio continuum
does not arise in AGN activity.  We investigate whether the 24\micron\ emission or
its morphological sub-components are distributed like the radio continuum and
molecular gas, which would strongly suggest star formation activity.  A schematic
summary is found in Table~\ref{schematic} with brief comments and additional
evidence from the FIR/radio flux density ratios, optical emission lines, 
stellar populations or GALEX colors where those are available.  
The FIR/radio flux density ratios are discussed in greater detail in
Section~\ref{radioFIR}.

\subsubsection{UGC 1503}

\begin{figure}
\centerline{
\includegraphics{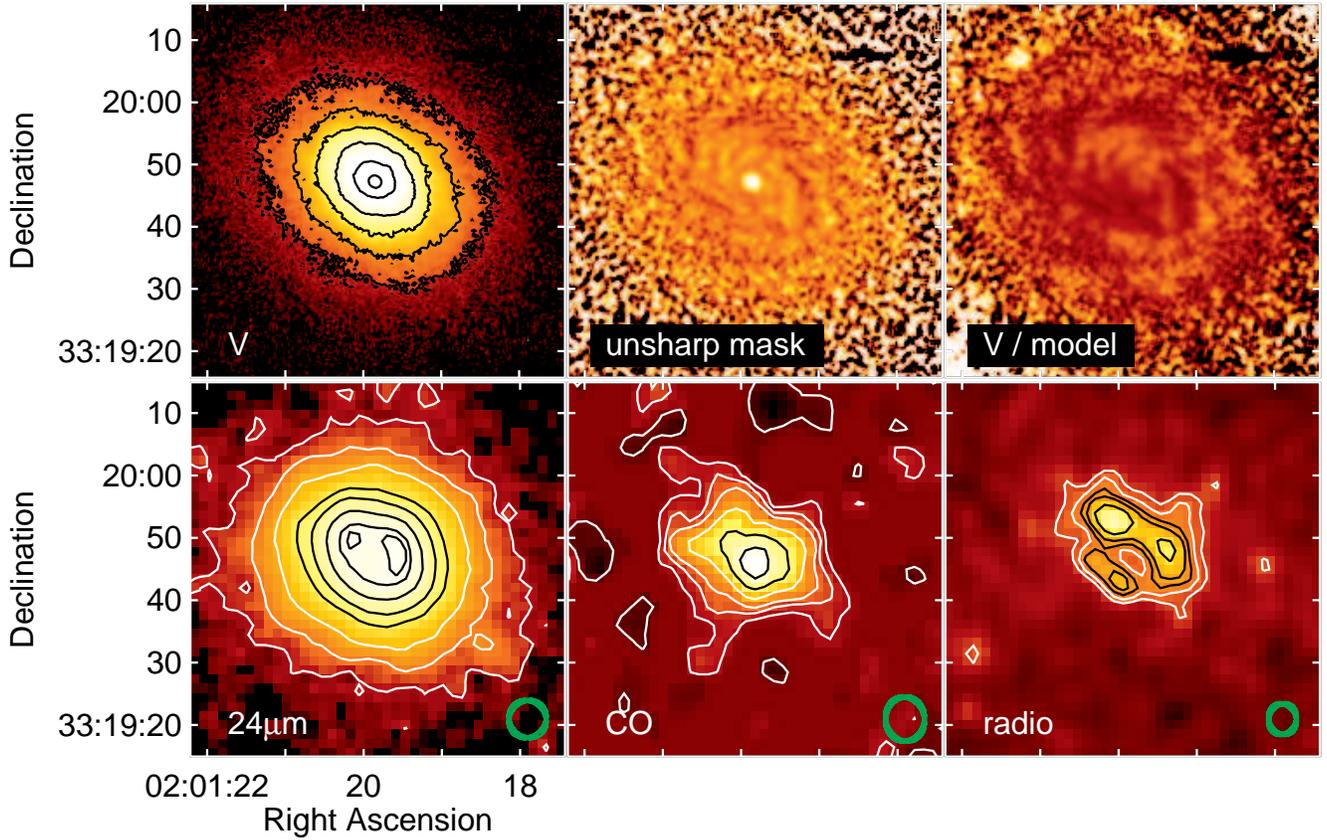}
}
\caption{\label{u1503} Optical, IR, CO and radio continuum morphology
of
UGC~1503.  Optical contours are spaced by a factor of two.  A $V-R$ image shows no
discernible structure but the unsharp masking and model division techniques do show some
faint structure.
Contour levels in the 24\mum\ image are 0.12, 0.31, 0.61, 1.22,
1.83, 3.05, 4.27, and 4.70 \Mjysr\ (the lowest three are 4.4$\sigma$,
11$\sigma$, and 22$\sigma$). 
Contour levels in the CO image are $-0.63$, 0.63, 1.27, 1.90, 3.12, 4.43, and
5.70 \jybkms. 
Contours in the 1.4 GHz radio
continuum image are 0.09, 0.12, 0.15, 0.18, and 0.24 \mjb\ (2.4, 3.1, 3.9, 4.7,
and 6.3$\sigma$).
In this and subsequent figures, green circles in the lower right corners of
panels indicate the resolution (FWHM) of their respective data.  Contours are
colored either white or black as necessary to enhance their visibility.
}
\end{figure}

Figure \ref{u1503} presents optical, unsharp-masked, 24\micron, CO,
and radio images of UGC~1503.  In addition, a smooth galaxy model was
constructed using the Multi-Gaussian Expansion technique of
\citet{MGE} and the top left panel of the figure shows the ratio of
the original $V$ image to the MGE model.  The unsharp-masked $V$ image
shows some fairly subtle mottling in the interior of the galaxy, with
perhaps some hint of flocculent arm segments at radii less than about
10\asec.  This is the same region where we find a regularly rotating
molecular gas disk (radius 15\asec\ = 5.2 kpc).  The radio continuum
emission is also found in an asymmetric ring of similar size.  The
24\micron\ image shows a central dip, which we have verified is not
due to saturation, and this image is well fit with a ring of radius
4.8\asec\ $\times$ 3.5\asec\ (1.6 $\times$ 1.2 kpc).  Table
\ref{fittable} gives the best-fit parameters describing the model of
UGC~1503 and the other galaxies.

A dust distribution following the stellar photospheric emission would
be centrally peaked, unlike the 24\micron\ emission of UGC~1503.  In
addition, if an annulus of dust in UGC~1503 were illuminated by an old
stellar population it could conceivably reproduce the 24\micron\
image, but as discussed above one would have to appeal to some other
process for the origin of the radio continuum ring.  For these reasons
we argue that the simplest explanation for the CO, radio and FIR
morphologies of UGC~1503 is that star formation is taking place in an
annulus of the molecular/dust disk.  The FIR/radio flux density ratios
support this interpretation, as discussed further in Section
\ref{discussion}.

While the global similarities between radio continuum and 24\micron\
morphology indicate that star formation activity in a $\sim$ 1.6 kpc
ring powers them both, the details of their morphologies have
implications for the propagation of the star formation through the
galaxy.  As Figure \ref{u1503} suggests, the ridgeline of the radio
continuum ring is clearly outside the ridgeline of the 24\micron\
ring, so that the two most prominent radio peaks are 3\asec\ to
4\asec\ away from corresponding 24\micron\ peaks.  The effect is not
caused by a difference in spatial resolution of the two bands; the
radio continuum image has somewhat better resolution (5\asec\ $\times$
4.5\asec\ $\sim$ 1.7 kpc) than the 24\micron\ image (6\asec), and the
absolute registration ought to be good to 1\asec\ or better.  Neither is this
the effect described in some detail by \citet{murphy08}, in which the
radio continuum emission looks like a smoothed version of the MIR or 
FIR image because the diffusion length of cosmic ray electrons is larger
than the mean free path of the photons that heat the dust. 
Instead, as the
radio continuum emission has a longer rise and decay timescale than
FIR emission after a burst of star formation activity
\citep{roussel03}, this offset could indicate that the star formation
activity is propagating inward or that the star formation rate is
decreasing more strongly in the outer parts of the ring than the inner
parts.

\subsubsection{NGC 807}\label{morph807}

\begin{figure}
\centerline{
\includegraphics{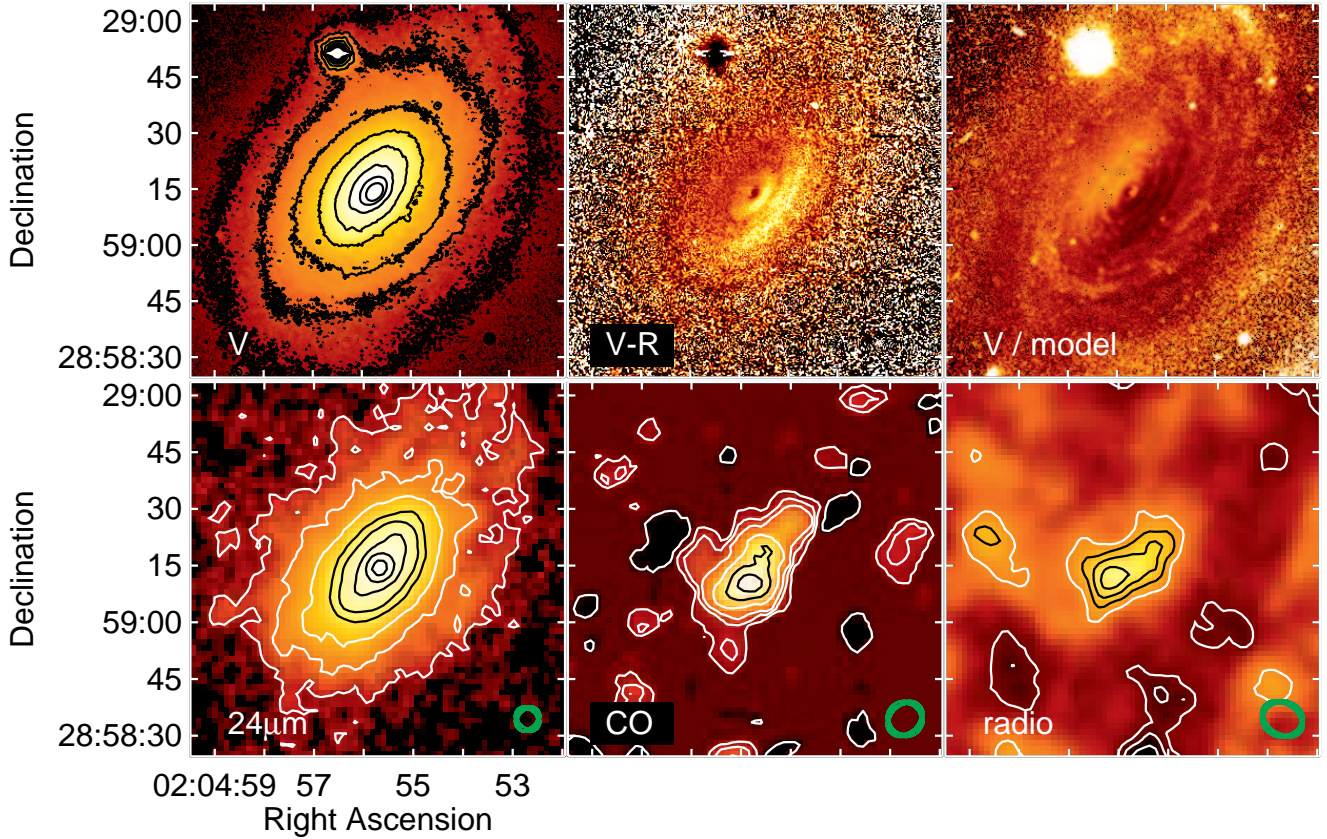}
}
\caption{\label{n807} Optical, IR, CO and radio continuum morphology
of
NGC~807.  Optical contours are spaced by a factor of two.  The top row,
middle
panel is a $V-R$ image and the top row, right panel is a $V$ image divided
by a
smooth MGE model constructed using the software of \citet{MGE}.
Contours in the 24\mum\ image are 0.12, 0.29, 0.58, 1.16, 1.74, 2.91, 4.07,
and 5.23 \Mjysr\ (the lowest three are 4.3$\sigma$, 10$\sigma$, and 20$\sigma$). 
Contours in the CO image are $-0.77$, 0.77, 1.53, 2.30, 3.83, 5.33, and 6.88
\jybkms. 
Contours in the 1.4 GHz radio
continuum image are $-0.2$, $-0.15$, $-0.1$, 0.1, 0.15, 0.2, and 0.25 \mjb\
($-3.6$, $-2.7$, $-1.8$, 1.8, 2.7, 3.6, and 4.5$\sigma$).
}
\end{figure}

Figure \ref{n807} shows the 24\micron\ emission from NGC~807 along
with an optical image, optical dust maps, CO line emission and 20 cm
radio continuum emission.  The dust distribution traced in the $V-R$
color map and in the MGE residuals suggests a symmetric, dynamically
relaxed disk.  The 24\micron\ emission from NGC~807 shows an
unresolved peak on the nucleus of the galaxy, an elongated plateau and
a surrounding envelope.  The structure of the emission is well fit by
a superposition of a nuclear point source (3.4 mJy), an exponential
disk of scale length 17\asec\ = 5.4 kpc, and a plateau or a
flat-topped inner disk whose flat region has a semimajor axis 11\asec\
= 3.5 kpc.  The two disks each contain roughly 50\% of the total
24\micron\ flux density and the point source only about 5\% of the
total.  At surface brightness levels of 1.2 \mjysr\ and higher (20\%
of the peak), the 24\micron\ emission shows a high degree of
reflection symmetry. That fact is significant because both the
molecular gas and radio continuum emission are notably stronger in the
southeast and their peaks are 6\asec\ southeast of the optical/24\mum\
nucleus.  (It is worth noting that the radio continuum emission from
NGC~807 is quite faint and data of higher sensitivity would be
beneficial.)

The nuclear point source component of the 24\micron\ image might be
related to either an AGN or to unresolved nuclear star formation.  It
is curious, though, that no corresponding nuclear peak is found in the
radio continuum emission.  The plateau or flat-topped disk component
is too extended to be attributed to an AGN, and as it does not follow
the stellar distribution either, it is unlikely to be the type of
circumstellar dust emission discussed by \citet{temi08}.  This
component could conceivably originate in star formation in the
molecular disk since its size scale is similar to that of the
molecular gas.  (The CO dist extends to radii $\approx$ 20\asec, and
the plateau is flat to a semimajor axis 11\asec\ and declines with a
scale length of 4\asec\ beyond.)  The third component in NGC~807, the
exponential component with a scale length of 17\asec, could also
conceivably originate in star formation in the molecular gas or it
could trace circumstellar emission.

If star formation activity strictly depends on the kpc-scale local gas
surface density, as in a Kennicutt-Schmidt relation \citep{kenn-m51},
we would expect it to be asymmetrically distributed like the molecular
gas.  But in the region where the molecular gas is found (radii less
than 20\asec) the 24\micron\ emission lacks this asymmetry, and this
observation suggests that either the 24\micron\ emission is not
primarily dust heated by star formation or that the local star formation
rate is determined by something else in addition to the local gas
density.  The radio continuum emission does seem consistent with star
formation in the molecular gas in both morphology and in total flux
density (Section \ref{discussion}).  In linear size this is quite a
large star-forming disk; the radius of the molecular disk is 20\asec\
= 6 kpc.  Thus, it seems likely that the 24\micron\ emission may arise
partly from star formation activity, but it must also include a
substantial contribution from more symmetrically distributed heat
sources such as an evolved stellar population and a nuclear source.

\begin{figure}
\includegraphics{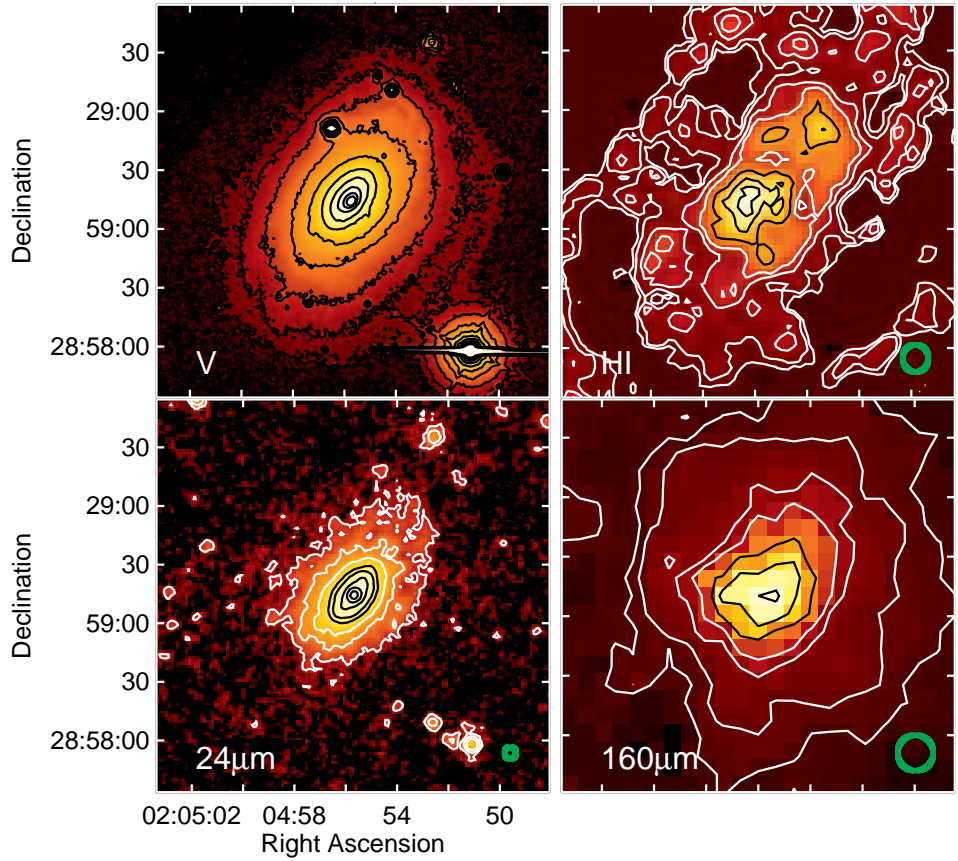}
\caption{\label{807-24-160} Large-scale optical, IR, and HI morphology of NGC~807.
Optical contours are spaced by a factor of two.  
Contour levels in the 24\mum\ image are the same as in Figure \ref{n807}.
Contour levels in the 160\micron\ image are 1.14, 2.28, 4.56, 6.84, 11.4,
16.0, and 20.5 \Mjysr\ (the lowest three are 2.0$\sigma$, 4.1$\sigma$, and
8.2$\sigma$). 
% rms there is 0.555 Mjysr.
The HI integrated intensity image has a
resolution of 14\asec\ $\times$
13\asec\ and contour levels are at 0.01, 0.02, 0.04, 0.06, 0.10, 0.14, and
0.18 \jybkms, where the peak in the HI intensity is 
0.20 \jybkms\ = 1.2\e{21} \persqcm.
}
\end{figure}

A unique feature of the 24\micron\ emission in NGC~807 is the very low
level, smooth emission extending to the west and northwest sides of
the galaxy at radii of 20\asec\ to 60\asec\ (6 to 19 kpc; Figures
\ref{n807} and \ref{807-24-160}).  This emission is at surface
brightness levels of 0.1 \mjysr\ to 0.3 \mjysr\ (3$\sigma$ to
10$\sigma$), and it is not accounted for in the parametric fit
described above.  Figure \ref{807-24-160} shows that the 160\micron\
emission is also extended to the northwest over the same spatial
region.  Similar asymmetries are found in the broadband optical and HI
images and are undoubtedly caused by a gravitational interaction that
has left large-scale disturbances in the outer parts of the galaxy.
For example, in the northeast and southwest corners of the HI image in
Figure \ref{807-24-160} there are sections of tidal arms that stretch
beyond the frame to radii of 5\arcmin\ (91 kpc).  HI column densities
northwest of the galaxy nucleus peak at 8.6\e{20} \persqcm\ (9.4
\msunsqpc, including helium) at a resolution of 14\asec\ (4.2 kpc);
this HI column density occurs 50\asec\ from the nucleus of the galaxy,
and is the brightest HI emission at radii $\geq 20''$ by a factor of
two.  Thus, the low level asymmetries in the 24\micron\ and
160\micron\ images of NGC~807 are probably consistent with an
interstellar dust component that is distributed like the atomic gas.

\subsubsection{NGC 2320}

\begin{figure}
\centerline{
\includegraphics{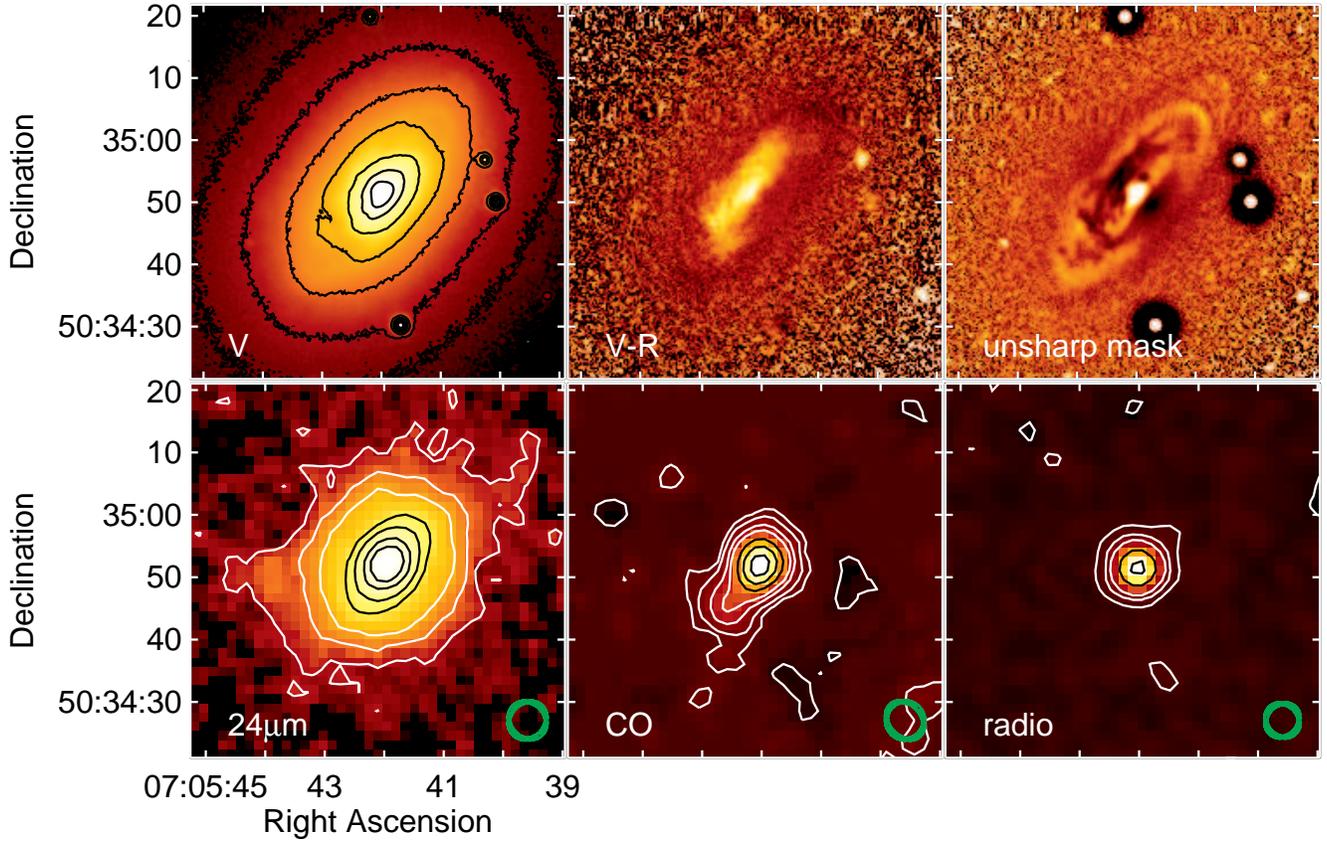}
}
\caption{\label{n2320} Optical, IR, CO and radio continuum morphology
of
NGC~2320.  Optical contours are spaced by a factor of two.  Contours in the
24\mum\ image are 0.11, 0.28, 0.56, 1.13, 1.69, 2.82, 3.94, and 5.08 \Mjysr\ 
(the lowest three are 3.9$\sigma$, 9.8$\sigma$, and 20$\sigma$).
Contours in the CO image are $-1.46$, 1.46, 2.91, 5.82, 8.73, 14.5, 20.4, and
26.2 \jybkms. 
Contours in the 1.4 GHz radio continuum
image are $-0.39$, 0.39, 1.31, 2.62, 6.55, and 11.8 \mjb\ ($-2.3$, 2.3, 7.7, 15,
39, and 69$\sigma$). 
}
\end{figure}

\citet{young05} showed that the $V$ morphology of NGC~2320 is well
described by an $r^{1/4}$ profile out to a semimajor axis of at least
90\asec, which bolsters its classification as an elliptical even
though it has quite a large molecular gas mass of 4\e{9} \solmass.
The $V-R$ image of Figure \ref{n2320} clearly shows the inner dust
disk to be 20\asec\ in diameter, with somewhat enhanced reddening
10\asec\ southeast of the nucleus.  The unsharp-masked image also
shows a bright ring of 35\asec\ diameter.  Molecular gas in NGC~2320
is distributed rather like the dust in the $V-R$ image and it also
shows a ``tail'' of gas to the southeast of the nucleus.  In contrast,
the radio continuum emission is not well resolved, with FWHM
$\lesssim$ 1.7\asec\ \citep{FIRST}, 
and it is believed to be powered by AGN activity as
described in greater detail by \citet{young05}.

Unlike the CO distribution in NGC~2320, the 24\micron\ emission is
notably symmetric.  It is well described by a $r^{1/4}$ model whose
half-light ellipse has semimajor and semiminor axes 4\asec\ $\times$
2\asec\ (1.6 $\times$ 0.8 kpc; Table \ref{fittable}).  The 24\micron\
$r^{1/4}$ model is much more compact than the optical emission from
the galaxy, though. Measured values for the effective or half-light
radius of this galaxy range from 13\asec\ ($J$, 2MASS Extended Source
Catalog) to 29.5\asec\ \citep[$V$,][]{cretton} to 37\asec\
\citep[$\sim B$,][]{RC3}, but all are significantly larger than our
fitted half-light radii at 24\micron.  We have confirmed the
discrepancy in effective radii by convolving the 2MASS $J$ image with
the 24\micron\ PSF and additionally by running our model fitting
software on the original 2MASS $J$ image to recover the effective
radius in the same way as was done for the 24\micron\ image.  We find
an effective radius of 23.4\asec\ \error\ 0.07\asec\ at $J$, more than
five times larger than the 24\micron\ effective radius.  In this
respect NGC~2320 is different from the ellipticals studied by
\citet{temi08}, which had very similar effective radii at optical/NIR
and 24\micron.  The difference in scale sizes implies that the
24\micron\ emission from NGC~2320 is not primarily circumstellar dust
around evolved stars, unless there is a strong age gradient in the
stellar population such that the inner arcseconds host a larger
proportion of intermediate-age stars that are relatively bright at
24\micron.

It is possible that the 24\micron\ emission in NGC~2320 is attributable to
star formation activity that is restricted to the central portions of
the molecular disk.  The 24\micron\ emission does not show the same
``tail'' to the southeast that is seen in the molecular gas, but these
might still be self-consistent if the molecular ``tail" is locally
gravitationally stable (perhaps not yet settled into its equilibrium orbit)
or if the local star formation rate has a
dependence on gas surface density that is steeper than linear.  The
CO column densities 10\asec\ southeast of the nucleus are at least a
factor of 3 lower than the peak in the center of the galaxy, and if
the star formation rate (traced by 24\micron\ emission) were
proportional to the square of the local molecular surface density then
the 24\micron\ emission 10\asec\ southeast of the nucleus would be a
factor of 10 fainter than the nuclear value.  But since the pointlike
radio continuum emission cannot be attributed to star formation
activity the current data do not necessarily favor star formation as
the heating source for the 24\micron\ emission.  More detailed
analysis of the ionized gas and the stellar population itself would
provide important circumstantial evidence about the origin of the
24\micron\ emission in this case.

\subsubsection{NGC 3032}

\begin{figure}
\centerline{
\includegraphics{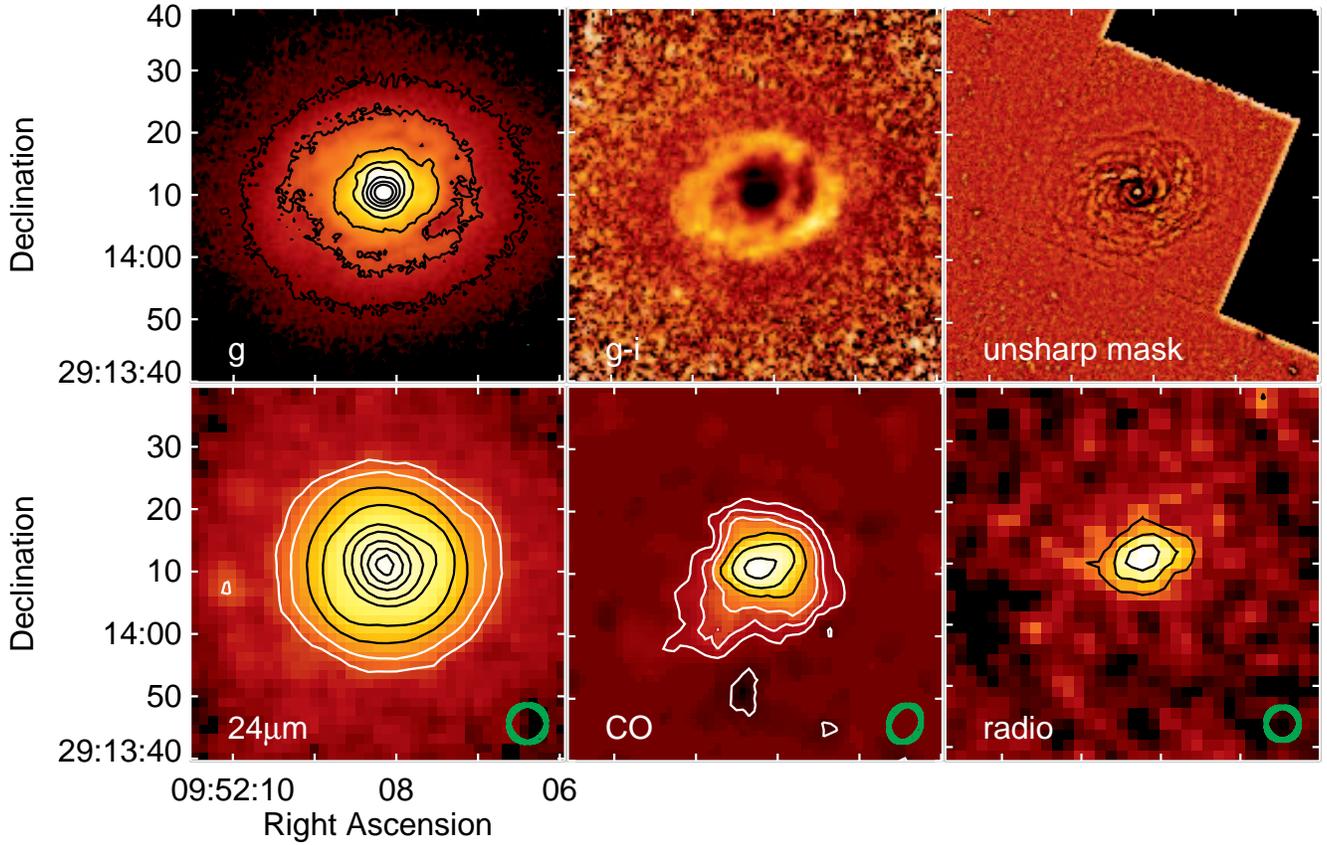}
}
\caption{\label{n3032} Optical, IR, CO and radio continuum morphology
of 
NGC~3032.  Optical contours are spaced by a factor of two. 
The unsharp-masked image is from a WFPC2 F606W image.
Contours in the
24\mum\ image are 0.51, 1.01, 2.53, 5.06, 10.1, 15.2, 25.3, 35.4, and 45.5
\Mjysr\ (the lowest three are 16$\sigma$, 32$\sigma$, and 81$\sigma$). 
Contours in the CO image are $-1.81$, 1.81, 3.62, 5.43, 9.05, 12.7, and 16.3
\jybkms. 
Contours in the 1.4 GHz radio continuum
image are 3.2, 5.4, and 7.5 times the rms noise level (0.14 \mjb).
}
\end{figure}

CO observations of this lenticular galaxy have recently been published
by \citet{ybc08}, and Figure \ref{n3032} shows the comparison between
optical, mid-IR, CO and radio morphology.  A clearly defined dusty
disk of diameter 28\asec\ is viewed at rather low inclination.  The
molecular disk is coincident with this dust disk, but it also has a
tail stretching to the southeast, and the tail has no obvious dust
counterpart.  Since the molecular gas is nearly exactly
counterrotating with respect to the stellar rotation \citep{ybc08}, it
is most likely to have been accreted from another galaxy or from the
intergalactic medium into its retrograde orbit, and the asymmetric
tail may be the remnant of the settling process.  Radio continuum
emission from NGC~3032 is very marginally resolved but is elongated in
the same direction as the dust and stellar disks.  The 24\micron\
emission from the galaxy is also very marginally resolved, being well
fit by an exponential disk of scale length 1.8\asec\ (0.19 kpc; Table
\ref{fittable}).  We find a marginal detection of an off-nuclear point
source that may or may not be physically related to NGC~3032.

Similarly to NGC~2320, and unlike the cases described in
\citet{temi08}, the 24\micron\ emission is a poor match to the
optical/NIR structure of the galaxy.  The 2MASS $J$ image of NGC~3032
was analyzed in the same way as we have done for the 24\micron\
images, and it was found best described by a model with 15\% of the
total luminosity in a compact nuclear source (probably a bright star
cluster) and 85\% in a $r^{1/4}$ spheroid with an effective semimajor
axis of 87\asec\ \error 2\asec.  Thus, again, the 24\micron\ emission
is not primarily circumstellar dust heated by evolved stars.

The east-west elongations of the dust disk, the radio continuum
emission and the bright part of the CO emission suggest that indeed
star formation activity could be powering the radio and 24\micron\
emission from NGC~3032.  Radio and FIR flux densities are also
consistent with star formation (Section \ref{discussion}), and there
is other evidence for star formation activity in the optical line
ratios \citep{sarzi06} and stellar Balmer line absorption
\citep{kuntschner06}.  Interestingly, the young counterrotating
stellar core found by \citet{mcdermid06a,mcdermid06b} has a radius of
roughly 2\asec, similar to the scale radius of the 24\micron\
emission, and this young stellar core is rotating in the same
direction as the molecular gas.  Thus the optical data are entirely
consistent with our interpretation that radio and 24\micron\ emission
in NGC~3032 arise in star formation.  This star formation activity is
most vigorous in the inner part of the molecular disk, which could be
due to a nonlinear dependence of the star formation rate on the local gas
density or to the gas disk being gravitationally stable in its outer regions.

\subsubsection{NGC 3656}\label{3656}

\begin{figure}
\centerline{
\includegraphics{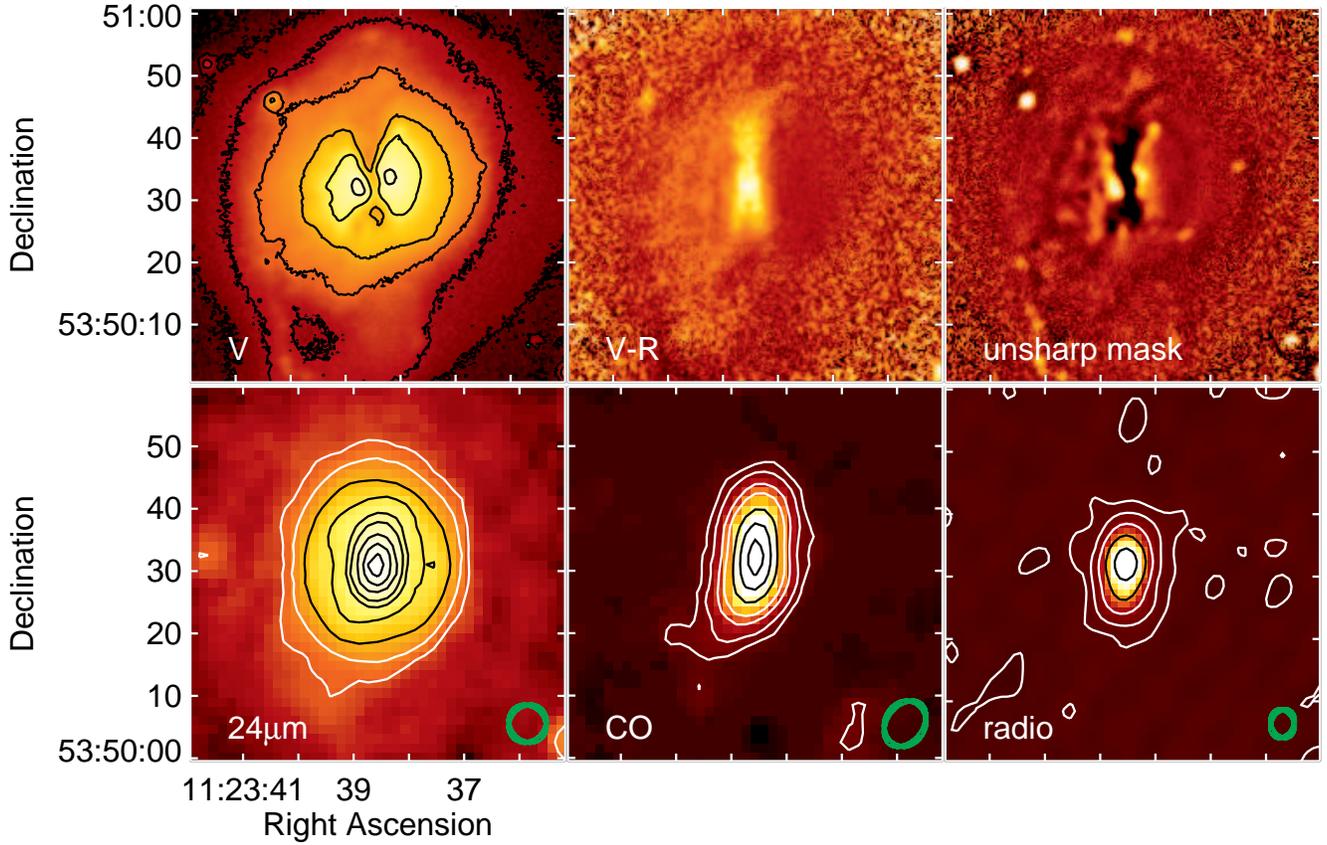}
}
\caption{\label{n3656} Optical ($V$), IR, CO and radio continuum morphology
of 
NGC~3656.  Optical contours are spaced by a factor of two. 
The unsharp-masked image is made from the $V$ image.
Contours in the
24\mum\ image are 0.57, 1.15, 2.87, 5.75, 11.5, 17.2, 28.7, 40.2, and 51.7
\Mjysr\ (the lowest two are 23$\sigma$ and 26$\sigma$). 
Contours in the CO image are 4.05, 8.09, 16.2, 24.3, 40.5, 56.6, and 72.8
\jybkms. 
Contours in the 1.4 GHz radio continuum
image are $-0.09$, 0.09, 0.3, 0.9, 2.4, and 6.0 \mjb\ ($-2.4$, 2.4, 8.1, 24, 64,
and 162$\sigma$).
}
\end{figure}

NGC 3656 is the most disturbed galaxy of this sample, being a clear
example of a fairly recent merger remnant with a dramatic blue ring
and shells.  Figure \ref{n3656} shows that the molecular gas is
located in a nearly edge-on disk of radius $\approx$ 10\asec\ = 1.9
kpc, also clearly visible as a dark dust lane bisecting the nucleus.
The structure of the dust lane is somewhat irregular, suggesting a
warped or folded lane rather than the very thin, flat, relaxed disks
seen in NGC 4526 and NGC 4459 (Sections \ref{4459} and \ref{4526}).
The radio continuum emission is clearly elongated in the same
direction as the gas/dust disk, with a deconvolved Gaussian source
size (HWHM) of 3.0\asec\ $\times$ 1.4\asec\ \citep{lucero}.  The 24
\micron\ emission is only modestly resolved but is also clearly
elongated north-south, and is adequately fit by an exponential disk of
scale length 2\asec\ \error 0.4\asec\ = 0.39 \error 0.08 kpc and axis
ratio 0.3:1.0 (Table \ref{fittable}).  Thus, the morphologies of radio
continuum, 24 \micron, and CO emission are all consistent with the
interpretation that star formation powers the bulk of the radio and 24
\micron\ emission.  The scale sizes of the radio and 24\micron\ disks
are substantially smaller than the molecular disk, as is also the case
for NGC~3032.  This result suggests that the local star formation rate
or efficiency is higher in the inner part of the gas disk.

\subsubsection{NGC 4459}\label{4459}

\begin{figure}
\centerline{
\includegraphics{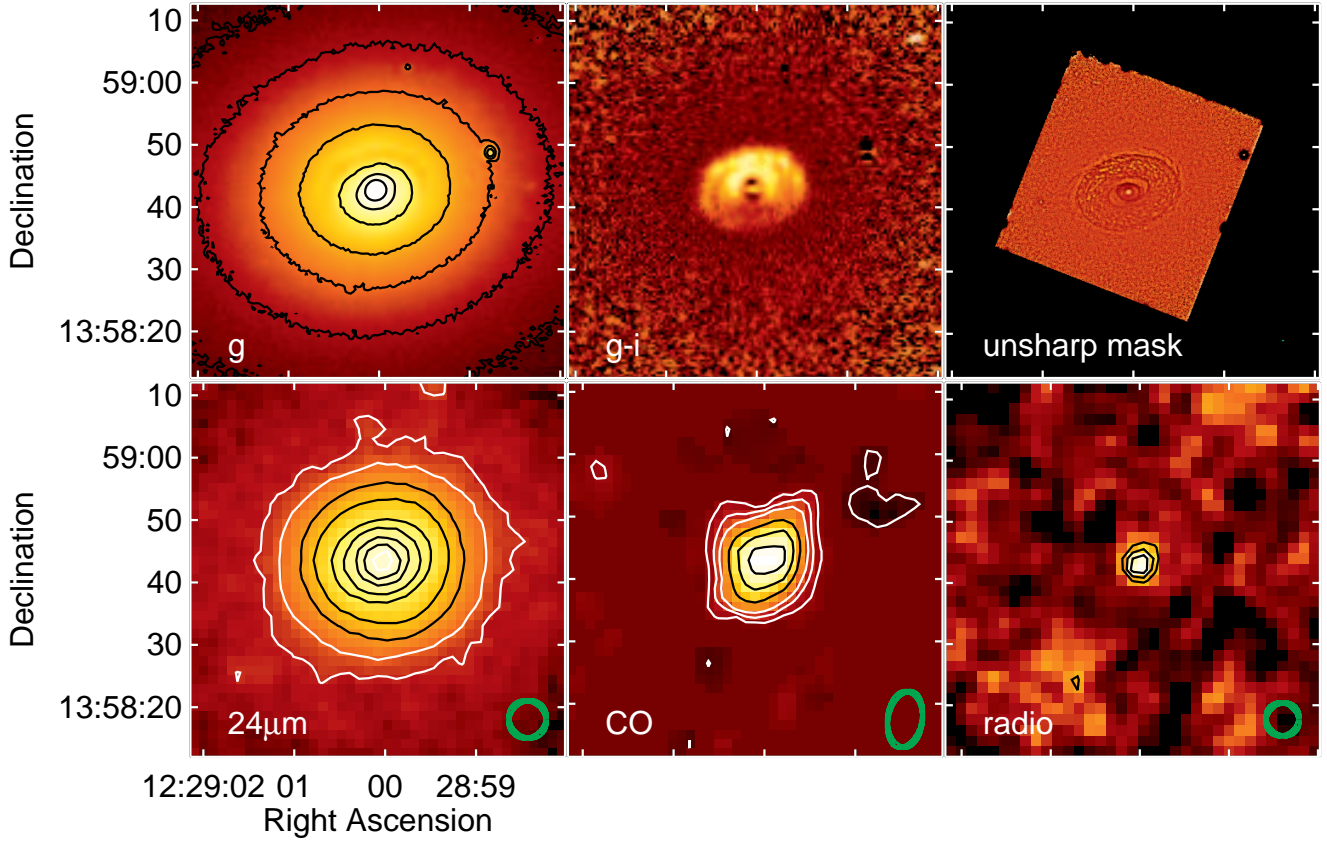}
}
\caption{\label{n4459} Optical, IR, CO and radio continuum morphology
of
NGC~4459.  Optical contours are spaced by a factor of two. 
Contours in the
24\mum\ image are 0.33, 0.66, 1.65, 3.30, 6.59, 9.89, 16.5, 23.1, and 29.7
\Mjysr\ (the lowest three are 8$\sigma$, 15$\sigma$, and 39$\sigma$). 
Contours in the CO image are $-2.14$, 2.14, 4.28, 6.42, 10.7, 15.0, and 19.3
\jybkms. 
Contours in the 1.4 GHz radio continuum
image are $-3$, 3, 4, and 5 times the rms noise level (0.15 \mjb).
}
\end{figure}

The Virgo Cluster lenticular NGC~4459 clearly shows a relaxed, thin
dust disk of semimajor axis 9\asec\ = 0.7 kpc (Figure \ref{n4459}),
and the unsharp-masked HST image resolves the dust into a broad,
flocculent outer annulus plus an inner ring of semimajor axis 2\asec.
The molecular gas lies in this regularly rotating disk, as shown in
the channel maps presented by \citet{ybc08}.  The radio continuum
emission from NGC~4459 is weak and unresolved-- in fact, it is
unusually weak, and is discussed further in Section \ref{discussion}.
The 24\micron\ emission is very symmetric and also centrally
concentrated but clearly resolved; it is well described by the sum of
two exponential disks of scale radii 2.1\asec\ (0.16 kpc) and 33\asec\
(2.6 kpc).  The more compact disk has 78\% of the total 24\micron\
flux density.

The more extended (33\asec) 24\micron\ component in NGC~4459 is
unlikely to be driven by star formation activity because its scale
length is so much larger than the size of the molecular gas and dust
disk.  It may in fact trace circumstellar dust in an evolved stellar
population as discussed by \citet{temi08}.  It is suggestive, though,
that the size of the more compact 24\micron\ component, 2\asec, is a
good match to the size of the inner dust ring in the top right panel
of Figure \ref{n4459}.  The radio continuum source is also of a
similar size.  Thus it is possible that the compact 24\micron\
component and the radio continuum trace star formation activity in the
central part of the gas disk (though we cannot rule out AGN activity
on morphological grounds alone).  Optical spectroscopy provides
abundant evidence for active star formation in NGC~4459; the SAURON
data show that the molecular disk coincides with an ionized gas disk
of low [O III]/H$\beta$ ratio \citep{sarzi06}, strong Balmer
absorption \citep{kuntschner06}, and a dynamically cold stellar
subpopulation \citep{emsellem04}.  If, then, 80\% of the 24\micron\
and all of the radio continuum in NGC~4459 trace star formation
activity, this is another case in which the efficiency must be much
higher in the inner part of the gas disk than in the outer part.

\subsubsection{NGC 4476}\label{4476}

\begin{figure}
\centerline{
\includegraphics{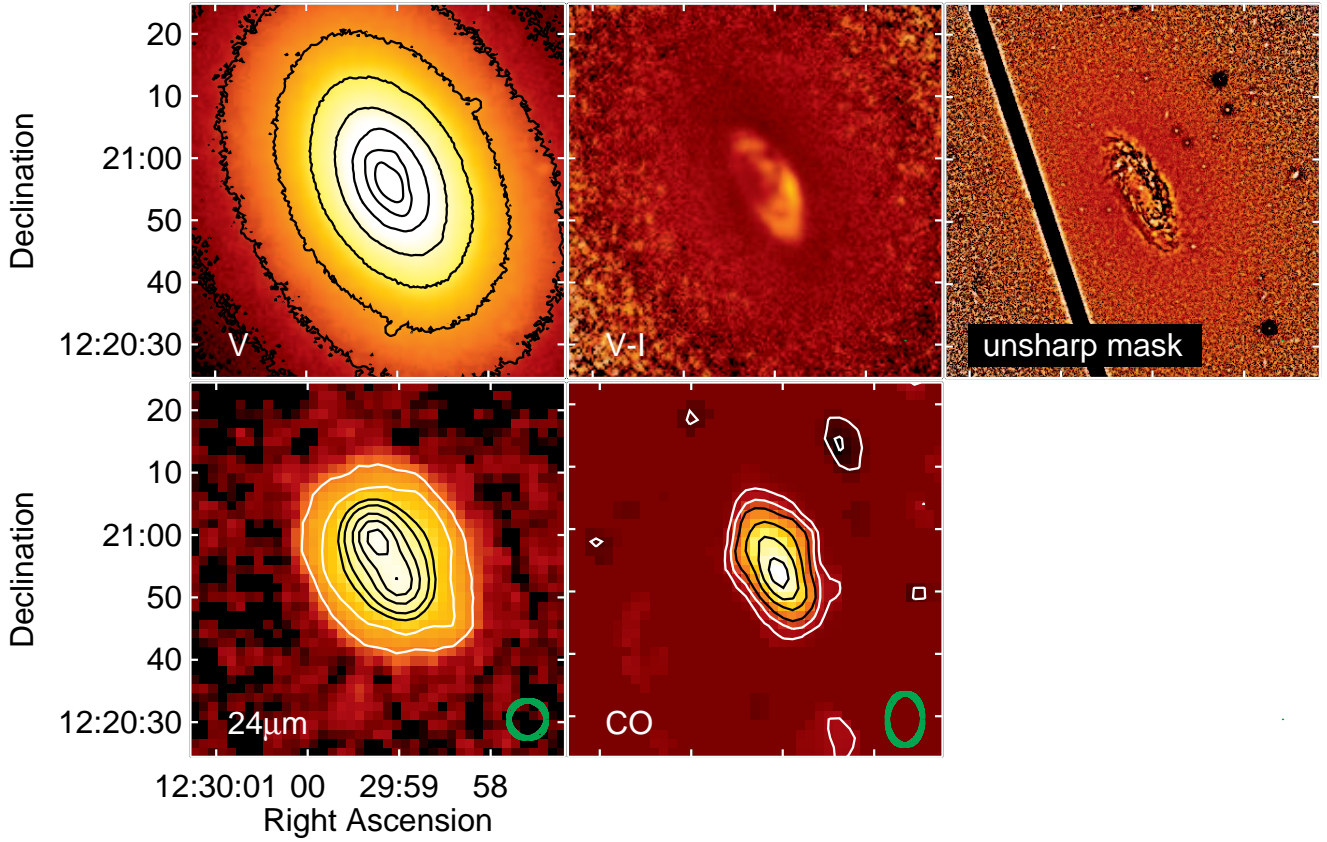}
}
\caption{\label{n4476} Optical, IR, CO and radio continuum morphology
of
NGC~4476.  Optical contours are spaced by a factor of two. 
Contours in the
24\mum\ image are 0.40, 0.80, 1.60, 2.40, 4.0, 5.6, and 7.2 \Mjysr\ (the lowest
three are 13$\sigma$, 25$\sigma$, and 50$\sigma$). 
Contours in the CO image are $-2.48$, $-1.24$, 1.24, 2.48, 3.72, 6.20, 8.68,
and 11.2 \jybkms. 
Radio continuum emission from NGC~4476 is
undetected \citep{lucero}.
}
\end{figure}

NGC~4476 is the second of our three Virgo Cluster early-type galaxies
with well-developed dust disks (Figure \ref{n4476}).   The F475W image also shows
numerous bright point sources embedded in the dust.  As for NGC~4459
and NGC~4526, the molecular gas is confined to this disk, which has a
semimajor axis of 11\asec\ = 0.9 kpc.  The bulk of the 24\micron\
emission can also be attributed to this dusty disk, as an excellent model of
the image is achieved from a ring with its maximum intensity at a
semimajor axis of 6\asec\ (0.5 kpc) and gradually decreasing
intensities on the interior and exterior (Table \ref{fittable}).  
Thus, there is
extremely good agreement between 24\micron\ morphology and CO
morphology.  

No radio continuum emission is detected from NGC~4476, and
\citet{lucero} hypothesize that the magnetic field and/or the
relativistic electrons may have been stripped by interactions with the
intracluster medium.  Circumstantial evidence for this idea comes from
the fact that no atomic gas is detected either \citep{4476hi}.  
We cannot, therefore, use the radio continuum morphology to argue that the
24\micron\ emission indicates star formation activity in this case.  
However, GALEX images of NGC~4476 show significantly bluer colors at radii
$\lesssim$~12\asec, becoming as blue as (FUV-NUV) $\sim$ 1.2 mag in the central
resolution element compared to (FUV-NUV) $\sim$ 2.5 mag in the outskirts of the galaxy
\citep{gdp07}.
In the absence of the radio continuum detection it is not
obvious that the 24\micron\ emission indicates star formation activity,
but the optical point sources and the GALEX colors strongly suggest this interpretation.

\subsubsection{NGC 4526}\label{4526}

\begin{figure}
\centerline{
\includegraphics{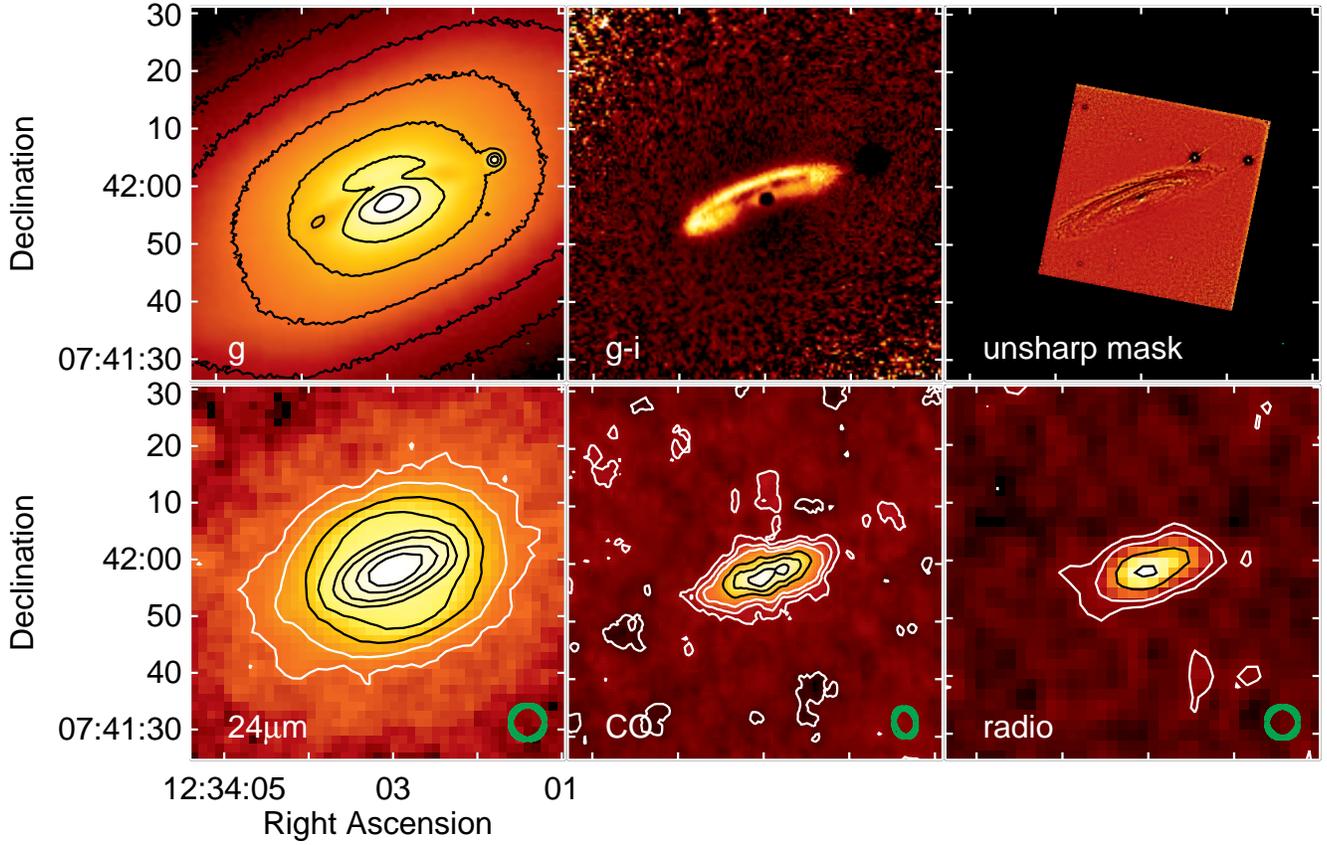}
}
\caption{\label{n4526} Optical, IR, CO and radio continuum morphology
of
NGC~4526.  Optical contours are spaced by a factor of two. 
Contours in the
24\mum\ image are 0.69, 1.38, 3.45, 6.90, 13.8, 20.7, 34.5, 48.3, and 62.1
\Mjysr\ (the lowest three are 7$\sigma$, 15$\sigma$, and 37$\sigma$). 
Contours in the CO image are $-3.71$, 3.71, 7.42, 11.1, 18.6, 26.0, and 33.4
\jybkms. 
Contours in the 1.4 GHz radio continuum
image are $-0.38$, 0.38, 0.75, 1.88, and 3.37 \mjb\ ($-2.5$, 2.5, 5.0, 12.5, and
22$\sigma$). 
}
\end{figure}

NGC~4526 is the third of our Virgo Cluster early-type galaxies with
well-developed molecular and dust disks (Figure \ref{n4526}).  The
gas/dust disk is seen at high inclination and has a sharp outer edge
at a semimajor axis of 15\asec\ = 1.2 kpc.  The radio continuum
morphology closely matches that of the molecular gas; the 24\micron\
emission is described by a plateau or a flat-topped disk of semimajor
axis 7.5\asec\ (0.6 kpc) and a more extended, rounder exponential disk
of scale length 28\asec\ (2.4 kpc).  The plateau contains 82\% of the
total flux density; given its size scale and orientation it is highly
likely that this 24\micron\ component is associated with the radio
continuum emission via star formation in the molecular gas.  The more
extended, fainter and rounder 24\micron\ component could well be
attributed to circumstellar dust as in \citet{temi07}, but it only
contributes 18\% of the 24\micron\ flux density.  Again, similar to
the case of NGC~4459, the optical spectroscopy traces star formation
activity and its after-effects in an ionized gas disk \citep{sarzi06}
and a young, dynamically cold stellar disk
\citep{kuntschner06,emsellem04} that have similar sizes to the
molecular disk.

\subsubsection{NGC 5666}

\begin{figure}
\centerline{
\includegraphics{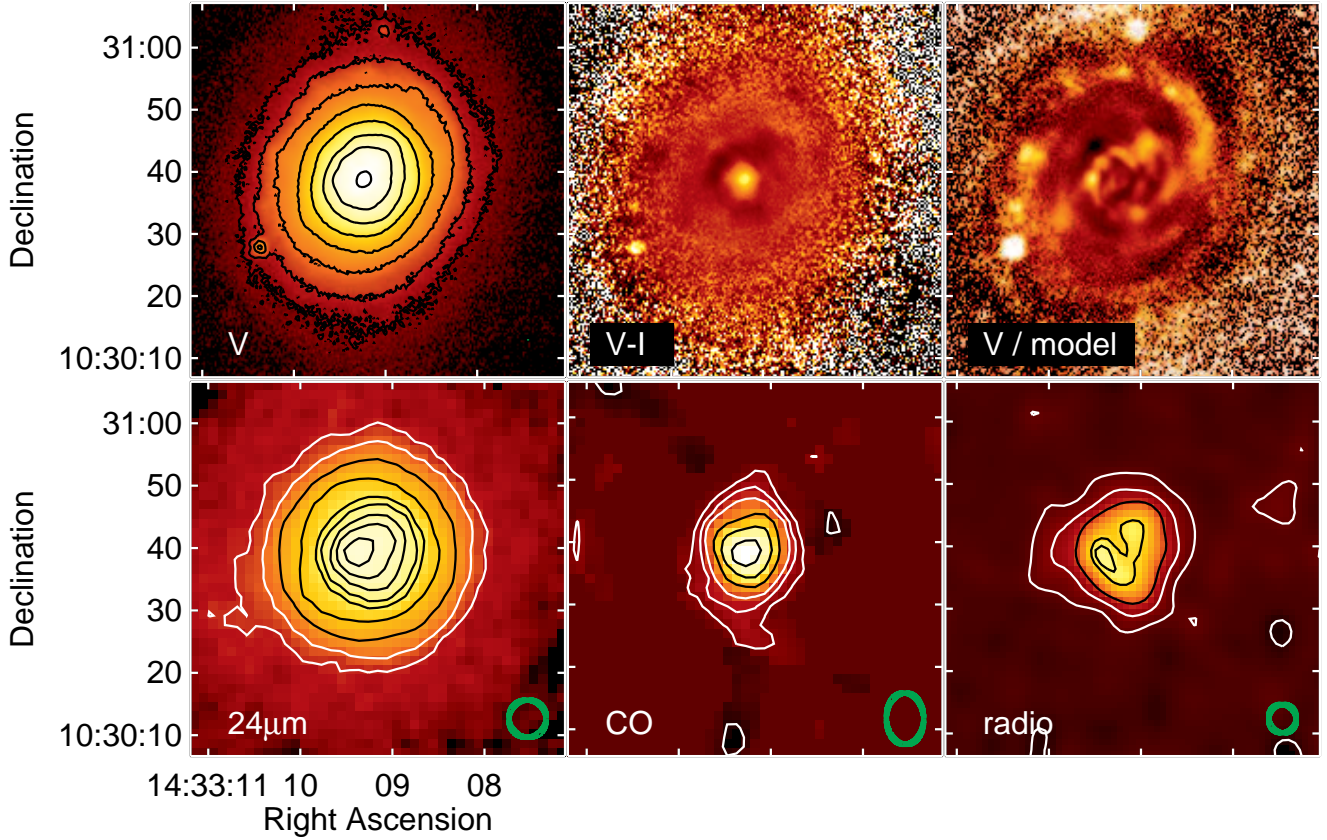}
}
\caption{\label{n5666} Optical, IR, CO and radio continuum morphology
of
NGC~5666.  Optical contours are spaced by a factor of two. 
The top right panel is constructed by dividing the $V$ image with a smooth 
elliptical MGE model.
Contours in the
24\mum\ image are 0.33, 0.67, 1.66, 3.33, 6.66, 9.99, 16.7, 23.3, and 30.0
\Mjysr\ (the lowest three are 12$\sigma$, 24$\sigma$, and 58$\sigma$). 
Contours in the CO image are $-1.68$, 1.68, 3.35, 5.03, 8.38, 11.7, and 15.1
\jybkms. 
Contours in the radio continuum image are $-0.12$, 0.12, 0.4, 0.8, 1.6, and 1.8
\mjb\ ($-2.4$, 2.4, 8, 16, 32, and 36$\sigma$).
}
\end{figure}

NGC~5666 was originally classified as an elliptical, but upon closer
inspection, its morphological status is uncertain.  Color images
(e.g.\ Figure \ref{n5666}) clearly show the red nucleus and a blue
ring of radius approximately 5\asec, which are also shown by
\citet{dd03} (DD03).  The ratio of the V image to its best-fit MGE
model also shows one spiral arm at radii of 15\asec -- 20\asec, and
DD03 find blue star-forming knots in the arm.  The arm would tend to
suggest a spiral classification but it is not obvious that the bulge +
exponential disk surface brightness fit of DD03 is well constrained or
is a better fit to the optical surface brightness profile than a
$r^{1/4}$ law.
In addition, DD03 comment that their bulge + disk decomposition has an
unusually extended bulge and compact disk, with the effective radius
of the bulge being larger than the scale length of the disk.  The
galaxy therefore does not seem a particularly close match to either a
prototypical elliptical or a prototypical spiral.  Measurements of the
stellar velocity dispersion and rotation velocity $(V_{max}/\sigma)$
would give better physical insight into the structure of this galaxy.

The molecular gas in NGC~5666 (Figure \ref{n5666}) is found in an
inclined disk of radius roughly 7--8\asec.  Radio continuum emission
is distributed in an asymmetric ring, brightest in the southeast and
the northwest, with a central dip.  Local maxima in the radio
continuum emission are at radii of 3.5\asec.  The best fit model to
the 24\micron\ image is a flat-topped disk or plateau of radius
5.9\asec = 0.9 kpc, but the outer scale length is so small that the
model is essentially an inclined top hat.  The observed 24\micron\
peak is offset a few arcseconds to the southeast of the galaxy nucleus
and it coincides with the peak of the radio continuum emission.

In this case the very close matches between the size of the 24\micron\
plateau, the extent of the molecular disk and the radio continuum ring
(as well as the blue ring in optical images) strongly suggest that
both the radio and 24\micron\ emission are driven by star formation
activity.  Models in which the 24\micron\ emission comes from a point
source or is distributed like the stars are strongly ruled out.

\subsection{70\micron\ and 160\micron\ Images} \label{70-160}

\begin{figure}
\includegraphics{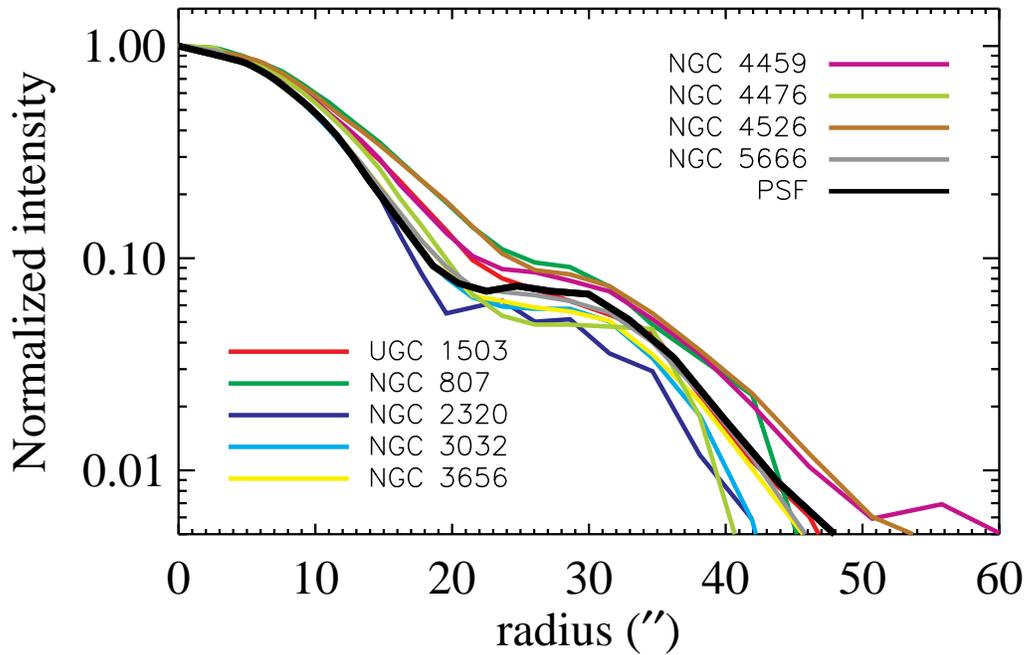}
\caption{Surface brightness profiles at 70\micron, normalized to 1.0 at the
center.  Uncertainties in the surface brightness due to the rms pixel-to-pixel
noise are at the level of a few percent near the first Airy minimum (20\asec) and
rise to 10\% in the radial range 40\asec\ to 60\asec.
\label{70SB}
}
\end{figure}

\begin{figure}
\includegraphics{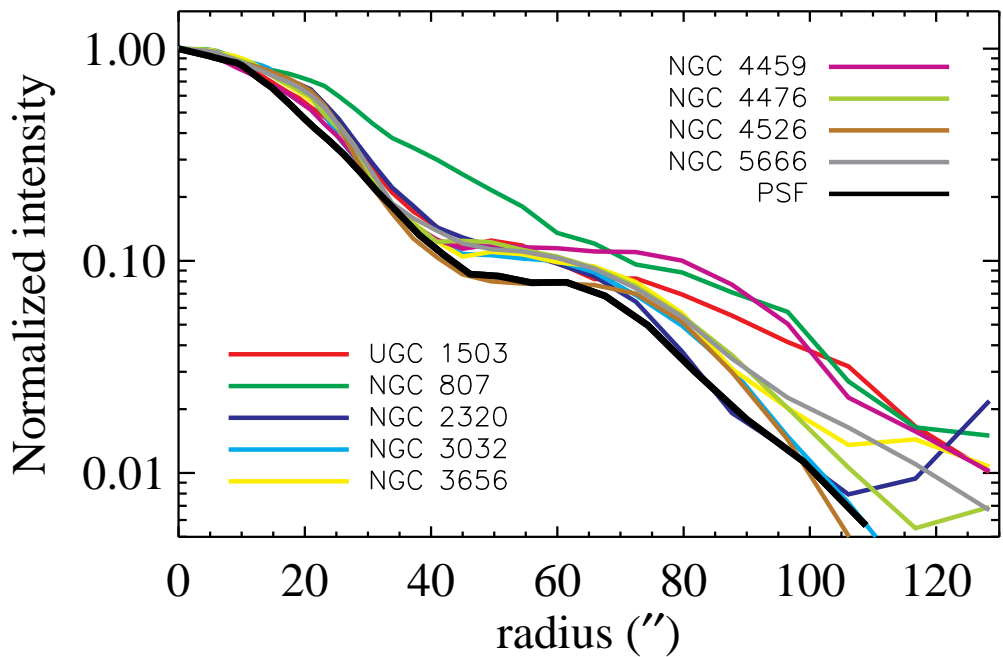}
\caption{Surface brightness profiles at 160\micron.  Uncertainties in the surface
brightness are a few percent for radii 20\asec\ to 40\asec\ and are $\gtrsim$
10\% at radii of 70\asec\ to 120\asec, depending on the surface brightness.
\label{160SB}
}
\end{figure}

The non-rotated images at 70\micron\ and 160\micron\ were fit with the
isophotal analysis program {\it ellipse} in STSDAS, and the resulting
surface brightness profiles are shown in Figures \ref{70SB} and
\ref{160SB}.  From these comparisons it is evident that, with the
exception of NGC~807, the targets are not resolved at 70\micron\ and
160\micron.  NGC~807 is modestly resolved at 160\micron\ and shows an
elongation in the same sense as the 24\micron\ image, as discussed
already in Section \ref{morph807}.

\section{Discussion}\label{discussion}

\subsection{Corroboration of Star Formation: Mid-IR vs Near-IR}

The 24\micron\ morphologies in these CO-rich early-type galaxies show
a variety of structures including disks, rings, point sources, and
very faint smooth emission.  We have argued that, at least in the
cases of UGC~1503, NGC~3032, NGC~3656,  NGC~4526, and
NGC~5666, and possibly also in NGC~4476, the bulk of the 24\micron\ emission must arise from dust
heated by star-forming regions.  In NGC~4526, we attribute the
24\micron\ emission from the elongated, flat-top disk component (82\%
of the total flux density) to star formation and the rest to
circumstellar emission.  In NGC~4459, we have argued that 78\% of the
24\micron\ emission is in a modestly-resolved component that is
probably heated by star formation though it could conceivably also be
powered by an active nucleus; the remaining 22\% we attribute to
circumstellar dust.  In NGC~807, approximately half to 95\% of the
24\micron\ emission may be related to star formation activity.  In
NGC~2320 the origin of the 24\micron\ emission is not clear,
so the most that can be said is that studies of the stellar
populations and the ionized gas distribution and line ratios would
provide useful clues.

Our morphological work also indicated that in our CO-rich early type
galaxies the 24\micron\ emission usually traces the molecular gas and
the silhouette dust disks more closely than it does the stellar
distribution.  In some cases this is fairly obvious, as the 24\micron\
emission is best modeled by a ring or a structure that is flat at
radii 6\asec\ to 11\asec.  In other cases the 24\micron\ model is
centrally peaked, but it has a radial scale length many times smaller
than that of the stellar distribution as we have verified by
analyzing near-IR $J$ images in the same way as the 24\micron\
images.

\begin{figure}
\includegraphics{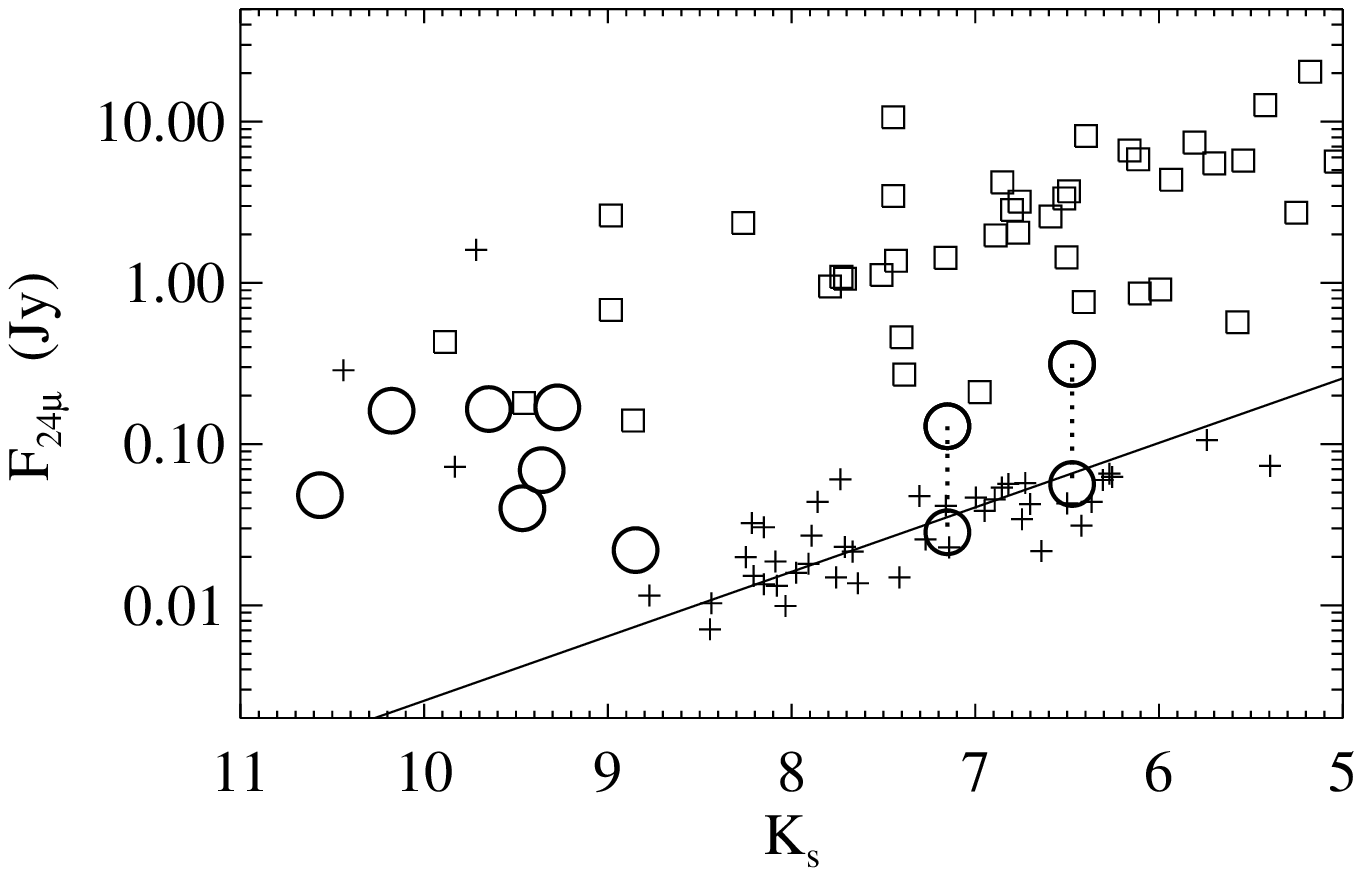}
\caption{\label{temifig} The 24\mum\ flux density
and $K_s$ apparent magnitude 
for the CO-rich early-type galaxies (circles).  For comparison, the sample of
ellipticals studied by \citet{temi07} are shown as crosses and spirals from the SINGS
sample are shown as squares \citep{dggetal07}.
In addition, for NGC~4459 and NGC~4526 we
show both the total 24\micron\ flux density and the more extended component by
itself, with dotted lines connecting the symbols.
Uncertainties are comparable to or smaller than the symbol sizes.
The solid line is not a fit but is instead a representative linear
relationship between the 24\micron\ and $K_s$ flux density.
In the generally CO-poor ellipticals of \citet{temi07}, 
the 24\mum\ emission is thought to be primarily
associated with mass loss from the evolved stars.  The extended 24\micron\ components in
NGC~4459 and NGC~4526 have 24\mum/$K_s$ ratios consistent with this behavior.
}
\end{figure}

This result is an obvious contrast to the findings of
\citet{temi07,temi08} that the 24\micron\ emission in their elliptical
galaxies closely traced the optical/NIR star light, with very similar
size scales and fairly tight correlations between $L_{24}$ and $L_B$.
However, that circumstellar dust should also be present in all of our
targets.  It must simply be outshone by the star formation.  As a test
of this hypothesis, Figure \ref{temifig} compares the 24\mum\ flux
densities and $K_s$ apparent magnitudes of our sample galaxies to
those of \citet{temi08}.  Integrated $K_s$ total magnitudes are taken
from the Two Micron All Sky Survey and its extended objects final release
as tabulated in the NASA Extragalactic
Database.  The median 24\micron\ /
2.2\micron\ flux density ratio in the CO-rich early-type galaxies is a
factor of 15 higher than that in the sample of \citet{temi08}; in
other words, the 24\micron\ flux densities are 15 times larger than
would be expected if the 24\mum\ emission were entirely circumstellar
in origin.  In Figure \ref{temifig} we also show the flux densities of
the extended 24\micron\ components of NGC~4459 and NGC~4526.  On the
basis of morphology we had previously argued that those components
would be consistent with the circumstellar emission; here we show that
their flux densities also support this interpretation because they
fall on the $24\micron - K_s$ correlation defined by the CO-poor
ellipticals.

\subsection{Additional Corroboration: FIR/Radio Ratios}\label{radioFIR}

\begin{figure}
\includegraphics{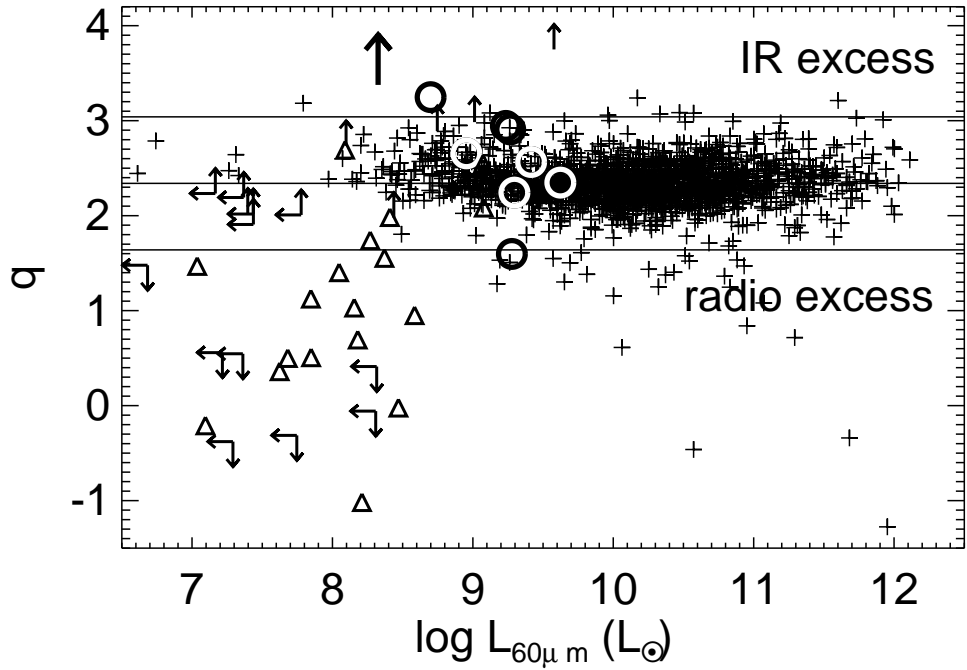}
\caption{FIR-to-radio flux density ratio $q$.  The circles and the large arrow
are the CO-rich
early-type galaxies studied here, and crosses are the data of \citet{YRC}.
As defined by those authors, the flux ratio $q$ has a mean value
of
2.34 for star-forming galaxies (mostly spirals); the lines at $q=3.04$ and
$q=1.64$ indicate the IR excess and radio excess boundaries, respectively, at 
roughly
$2.7\sigma$ from the mean.  
Uncertainties are comparable to or smaller than the symbol sizes, though we have
not included the effect of the distance uncertainty on the 60\micron\ luminosity.
The ratio $q$ is, of course, independent of distance.
Evidently the CO-rich early-type galaxies follow the
same radio/FIR relation as the star forming spirals. Another sample of 48
elliptical galaxies from \citet{temi07} are shown in triangles and small arrows.
\label{qplot}
}
\end{figure}

Radio to FIR flux density ratios confirm the suggestions of star
formation activity in all cases except NGC~2320.  For the computation
of these ratios we follow \citet{YRC} (YRC), who have used IRAS
60\micron\ and 100\micron\ flux densities; the choice is made because
YRC provide the most complete comparison set of FIR and radio
continuum flux densities for many types of galaxies.  The logarithmic
FIR/radio ratio $q$ is presented in Figure \ref{qplot} for our sample
of CO-rich early type galaxies as well as for YRC's sample (which
includes all galaxies having 100\micron\ flux densities greater than 2
Jy) and the early-type galaxies of \citet{temi07}.

According to the definitions of YRC, NGC~2320 is a radio excess galaxy
whose radio continuum is likely powered by an AGN; this classification
is consistent with its point source radio morphology.  NGC~4476 and
NGC~4459 are FIR-excess galaxies.  The remainder of the galaxies have
FIR/radio ratios $q$ that are consistent with those of star forming
galaxies and that confirm our inferences based on the 24\micron\
morphology.  For comparison, the majority of the elliptical galaxies
studied by \citet{temi07} are radio excess galaxies hosting AGN
emission, though some of them do also sit near the star formation
FIR-radio correlation.

The FIR-excess galaxies are rare, comprising 9 of 1809 galaxies in the
sample of YRC, although they are somewhat more common in other samples
of spirals \citep{roussel03}.  Being rare, they are valuable for the
insights they give into the physical processes underlying the
FIR/radio correlation.  For example, both YRC and \citet{roussel03}
note that some of the FIR-excess galaxies have unusually high dust
temperatures as indicated by IRAS 60\mum/100\mum\ flux density ratios
$\ge 0.78$.  Others have ``normal'' dust temperatures.  YRC have
argued that the ``hot dust'' FIR excess galaxies may be very densely
enshrouded AGN or compact nuclear starbursts.
\citet{roussel03,roussel06} have argued that a few of the hot dust FIR
excess galaxies are likely to be cases of recently initiated
starbursts, in which the starburst began so recently that there has
not yet been time for appreciable numbers of supernovae to accelerate
the cosmic ray population and produce the cm-wave synchrotron
emission.  They have also argued that ``cool dust'' FIR-excess
galaxies are dominated by heating from the evolved stellar population,
not star formation \citep[c.f.][]{temi08}.

All of the CO-rich targets studied here fall into the ``cool dust''
category, and our FIR excess galaxies NGC~4476 and NGC~4459 have IRAS
60/100\mum\ flux density ratios of 0.36 and 0.39 respectively.  Thus,
of the three preceding explanations for the FIR excess phenomenon, we
reject the idea of highly enshrouded AGN in our targets.  We have also
used morphology to argue that the bulk of the 24\micron\ emission does
not originate from dust heated by the evolved stellar population in
NGC~4459.  It is possible that recently initiated star formation
could be responsible for the relatively low level of radio continuum emission in
NGC~4476 and NGC~4459 (and perhaps also in NGC~0807 and NGC~4526,
which have relatively high $q$ values).  
Close inspection of their stellar populations could test that idea.
An alternative, possibly viable explanation is that the weak radio
continuum flux densities may be related to unusually weak magnetic
fields.  In the case of the Virgo cluster members NGC~4459, NGC~4476,
and NGC~4526, the interactions with the intracluster medium could be
responsible for stripping the plasma and/or the magnetic fields in a
more advanced version of the stripping that is now evident in the
Virgo spirals studied by \citet{murphy-conf}.  High quality radio
polarization imaging would help to answer this question, and such work
would also be valuable because rather little is known about the
strengths of interstellar magnetic fields in early-type galaxies.

\cite{draine07} have made careful models of the mid- to far-IR emission  in
nearby spirals and have concluded that the FIR emission (specifically,
emission at wavelengths longer than 100\micron) originates in the diffuse ISM,
not star forming regions. 
If this is true, then the FIR emission in spirals may be a better tracer of the total
dust mass than of the total light emitted from star-forming regions.
FIR emission in spirals may still be associated with star formation, but as a cause
rather than an effect; the association could be through the Schmidt law in the same
way that molecular clouds are associated with star formation.
In this case, the tightness of the
radio-FIR correlation is still puzzling.  Galaxies that are rich in cold
neutral gas but are not forming stars (such as NGC 2320?) 
should be particularly valuable
test cases for disentangling the physical processes driving these
emissions, and it would be helpful to have FIR images at spatial
resolutions matching those of the molecular gas maps.
Aside from the slight extension in the 160\micron\ image of NGC~807 there
is still very little information on the resolved FIR emission of early-type
galaxies.  One notable exception is that the 
ground-based observations of \citet{leeuw08} have
clearly resolved the 350\micron\ emission in at least UGC~1503, NGC~807,
NGC~3656, and NGC~4476, and they show it to be 
elongated at the same orientation as our 24\micron\ and CO
images.  

\subsection{The Future and the History of these Galaxies}

\citet{calzetti} find an empirical relation between the 24\micron\
luminosity of a star-forming region or galaxy and its star formation rate:
$$ \mathrm{SFR}\; (M_\odot\; yr^{-1}) = 1.27\times 10^{-38}\; (L(24 \micron))^{0.885}.$$
Here $L(24 \micron)$ is the monochromatic luminosity at 24\micron\ in units of
erg~s$^{-1}$, given by 
$4\pi D^2\, \nu S_{\nu}$ with 
$S_\nu$ being the flux density at 24\micron\ and $\nu$ 
the central frequency of the 24\micron\ band.  If this relation is accurate in
early-type galaxies as well (e.g.\ if the initial mass function is the same in
both types of galaxies, among other issues), 
the star formation rates in our CO-rich targets are on the
order of a few tenths of a solar mass per year (Table \ref{SFRs}).  Gas
depletion timescales can be estimated from the total molecular gas mass 
(molecular hydrogen with a correction for helium) divided by the star formation
rate.  Those timescales are a few Gyr, with one
exception.  The 24\micron\ luminosity of NGC~2320 is so low compared to its
total molecular mass that even if all of its 24\micron\ luminosity were powered
by star formation the gas depletion timescale would be longer than the Hubble time.

Since the molecular gas in the early-type galaxies settles into kpc-scale disks
\citep{young02,young05,ybc08}, the young stars will also be dynamically cold
populations.  The CO-rich early-type galaxies,  with the exceptions already
discussed, are building disky stellar components on timescales of a few Gyr.  
Indeed, in
NGC~4526 the dynamically cold stellar disk is already visible in the stellar
kinematics \citep{emsellem04}.  However,
since the total molecular gas masses are only a few \tenup{8}~\solmass\ to a few
\tenup{9}~\solmass, the young stellar populations will not make up more than a
few percent of the total stellar mass.

The gas depletion timescales in Table \ref{SFRs} are similar to the
corresponding timescales in the disks of spiral galaxies.  Depletion
timescales are, of course, very sensitive to assumptions about initial mass
functions and CO-to-H$_2$ conversion factors, but estimates for spirals tend to
be in the range of a few Gyr \citep{kenn-araa} and the
values in Table \ref{SFRs} are 1 to 4 Gyr in seven of nine cases.
In other words, while the elliptical and
lenticular galaxies have smaller molecular masses than spirals, their star formation
efficiencies are similar to within factors of a few.   

\citet{kenn-araa} has also pointed out that recycled
matter from stellar evolution (especially from the more massive stars) may make
the gas reservoir last a factor of 2 to 3 times longer than the simple depletion
timescale estimate.  Thus, we can expect the molecular disks in the early-type
galaxies to remain for many Gyr, perhaps as long as another Hubble time, unless
there are significant destructive processes such as stripping, interactions with
hot gas, and/or possible AGN feedback.  Ram pressure stripping will be more intense
in clusters than in the field, and tidal interactions more frequent, so the 
environments of our CO-rich early-type galaxies are also of interest.  
NGC~4476, NGC~4459, and NGC~4526 are members of the Virgo Cluster and NGC~2320 is a
member of the Abell 569 cluster; NGC~4476 is already known to have suffered some ram
pressure stripping \citep{4476hi}, but it still retains its molecular gas at least 
for the time being.

To put this disk growth into the broader context of galaxy formation
and evolution, it is important to know what (if anything) the molecular content
of an early-type galaxy reveals about its history.  
However, the origins of the molecular gas in early-type galaxies are not yet
known in most cases, and it is not yet clear why some early-type galaxies are rich
in molecular gas while others are not.
For many years there has been speculation that the molecular gas could originate 
from the mass loss of evolved stars \citep[e.g.][and references therein]{sw06}.
In other cases such as NGC~3032 \citep{ybc08} and NGC~2768 \citep{crocker08}, the
clear kinematic misalignments between molecular gas and stars indicate that the
molecular gas was accreted from some external source or possibly it is leftover
from a major merger.  It will be necessary to obtain larger numbers of
molecular gas maps and better statistics on the CO content of the early-type
galaxies before we will understand what fraction of their molecular gas
could have come from internal and external origins,
or whether the CO-rich and the CO-poor galaxies have had systematically
different formation histories.

\section{Summary}

For the majority of the CO-rich early-type galaxies, the close
agreements between CO, 24\mum, and radio continuum morphologies
suggest that the bulk of the 24\mum\ emission should be attributed to
star formation activity.  In our sample, these galaxies include
UGC~1503, NGC~3032, NGC~3656, NGC~4459, NGC~4526, and
NGC~5666.  The 24\micron\ emission in NGC~4476 is likely due to star
formation.  A portion of the 24\micron\ emission from NGC~807 may also
trace star formation activity, but the origin of the 24\micron\
emission in NGC~2320 is not yet clear.  Radio and FIR flux density
ratios are consistent with this interpretation, as are the increased
${24\mu m}/K_s$ flux denstiy ratios of the CO-rich over the CO-poor
early-type galaxies.  Thus, one of the major implications of this work
is that the CO, radio, and MIPS data are roughly consistent with the
UV results implying star formation activity in a few tens of percent
of the nearby early-type galaxies.  The necessary raw material is
present, more or less, and the molecular gas is often being processed
into stars.

In the cases of NGC~3032, NGC~3656, and NGC~4459, the star
formation activity seems to be taking place on more compact spatial
scales than the distribution of the gas itself.  This situation could
arise if the star formation rate is a strong (nonlinear) function of
the local gas surface density, or if the molecular disks are only
unstable to star formation in their inner portions.

However, the origin of the 24\micron\ emission in the CO-rich
early-type galaxies is not entirely clear-cut.  An extended component
in NGC~807 (detected both at 24\micron\ and at 160\micron) may trace
dust in the diffuse atomic ISM.  Extended 24\micron\ components in
NGC~4459 and NGC~4526 may arise from dust around the evolved stars,
just as \citet{temi08} find in the CO-poor early-type galaxies.
Likewise, though NGC~2320 is very CO-rich, the mismatches between CO,
radio continuum and 24\micron\ morphology make it unclear how much (if
any) star formation is occurring there.  In these cases, detailed
comparisons with the UV morphologies and the stellar populations would
provide firmer information about the distribution of star formation
activity (or the lack of it) within the galaxies' molecular disks.

This work also opens the way for more quantitative tests of
theoretical and phenomenological models of the star formation process.
For example, with a model of the gravitational potential (from the
stellar distribution and a mass-to-light ratio), it would be possible
to test whether the Toomre-type local gravitational instability is
consistent with the locations of star formation activity, as discussed
by Kawata et al.\ (2007).  The gas-rich early-type galaxies could also
provide useful perspective on the workings of the Kennicutt-Schmidt
relationship between the star formation rate and the gas surface
density.  If these models have truly captured some underlying physics
of the star formation process they ought to work in the ellipticals
and lenticulars as well as in the spirals.

\acknowledgments

LMY thanks the University of Oxford sub-department of Astrophysics and the
Imperial College department of Astrophysics for their hospitality.
Scott Montgomery provided assistance in the early stages of
the project.
We also thank Liese van Zee for the WIYN images and Daniel A.\ Dale for the use of his SED template and for discussions
about his results.
This research has made extensive use of the NASA/IPAC Extragalactic Database (NED)
which is operated by the Jet Propulsion Laboratory, California Institute of
Technology, under contract with the National Aeronautics and Space
Administration.
This work is based on observations made with the {\it Spitzer Space
Telescope}, which is operated by the Jet Propulsion Laboratory (JPL),
California Institute of Technology under NASA contract 1407.
Support for this work was provided by NASA and through JPL Contract 1277572.

\begin{deluxetable}{lrrrcrccccl}
\tablewidth{0pt}
\tabletypesize{\footnotesize}
\tablecaption{Sample Galaxies -- Basic Properties \label{sampletable}}
\tablehead{
\colhead{Name} & \colhead{RA} & \colhead{Dec} & \colhead{Dist} &
\colhead{Type} & \colhead{$M_B$} & \colhead{$(B-V)_e$} & 
\colhead{$R_e$} & \colhead{$\sigma_0$} & \colhead{M(H$_2$)} & \colhead{Refs.} \\
\colhead{} & \multicolumn{2}{c}{(J2000.0)} & \colhead{Mpc} & \colhead{} &
\colhead{} & \colhead{} & \colhead{\asec} & \colhead{\kms} &
\colhead{$10^8$ M$_\odot$} & \colhead{} }
\startdata
UGC 1503 & 02 01 19.8 & +33 19 46 & 71 (5) & E & $-20.1$ & \nodata     & 8   &
\nodata & 19 (3) & 8 \\
NGC 0807 & 02 04 55.7 & +29 59 15 & 66 (5) & E & $-20.8$ & 0.97 & 12   &
175 & 14 (3) & 8\\
NGC 2320 & 07 05 42.0 & +50 34 42 & 84 (7) & E & $-21.7$ & 1.05 & 13 &
350 & 49 (7) & 1, 2, 3\\
NGC 3032 & 09 52 08.2 & +29 14 10 & 21.2 (1.9)  & S0$^0$ &  $-18.8$ &
0.63 & 12 & 82 & 4.9 (1.0) & 4, 5\\ % SAB0^0(r)
NGC 3656 & 11 23 38.4 & +53 50 31 & 40 (3) & I0pec &  $-19.8$ & \nodata
& 14   &  180   & 37 (5) & 8\\
NGC 4459 & 12 29 00.0 & +13 58 43 & 16.1 (0.4) & S0$^+$ & $-20.0$ &
0.97 & 25 & 174 & 1.7 (0.3) & 6\\ % S0^+(r)
NGC 4476 & 12 29 59.2 & +12 20 55 & 17.6 (0.6) & S0$^-$ & $-18.3$ &
0.85 & 10   & 53  & 1.0 (0.1) & 6\\ % SA0^-(r)
NGC 4526 & 12 34 03.0 & +07 41 57 & 17.3 (1.5)  & S0$^0$ & $-20.8$ &
0.98 & 44 & 256 & 6.3 (1.1) & 2, 4, 7\\ % SAB0^0(s):
NGC 5666 & 14 33 09.1 & +10 30 37 & 31 (2) & ? & $-18.7$ & 0.86 &
 6 & \nodata  & 4.5 (0.6) & 8\\
\enddata
\tablecomments{
Distance references - (1) \cite{rstetal04}; (2) \citet{rtss05}; (3)
\citet{prs06}; (4) \citet{tonry01}; (5) \citet{jtbtlrab03}; (6)
\citet{metal07}; (7) \citet{wwpzl06}; (8) \citet{RC3}.  In the latter case,
distances are calculated using the velocity relative
to the galactic center of rest and $H_0=73
\pm 5$~km~s$^{-1}$~Mpc~$^{-1}$.
The CO fluxes are taken from \citet{young02, young05} and \citet{ybc08}, 
and H$_2$ masses use a
CO-to-H$_2$ conversion factor of 3.0\e{20} \convunits\ at the updated
distances.
Effective radii are the $J$-band half-light radii from the 2MASS extended source
catalog; note that in the cases of NGC~2320, NGC~4459, and NGC~4476 those values
are significantly lower than the effective radii quoted  by \citet{RC3}, which
are 37\asec, 35\asec, and 18\asec\ respectively.
Other data are taken from NASA's Extragalactic Database (NED) and the
Lyon Extragalactic Database (LEDA).  The morphological type of NGC~5666 is
uncertain, as discussed in the text.
}
\end{deluxetable}

\begin{deluxetable}{lccccc}
\tablewidth{0pt}
\tablecaption{Radio Continuum Image Parameters
\label{radioims}}
\tablehead{
\colhead{Galaxy} & \colhead{Beam} & \colhead{noise} &
\colhead{Peak} & \colhead{S/N} & \colhead{Ref.} \\
\colhead{} & \colhead{\asec} & \colhead{\mjb} & \colhead{\mjb} &
\colhead{} & \colhead{}
}
\startdata
UGC 1503 & 5.0$\times$4.5 & 0.038 & 0.33 & 8.7 & LY07\\
NGC 0807 & 10.9$\times$8.8 & 0.056 & 0.27 & 4.8 & LY07\\
NGC 2320  & 5.4$\times$5.4 & 0.17 & 13.1 & 77 & FIRST\\
NGC 3032  & 5.4$\times$5.4 & 0.14 & 1.22 & 8.7 & FIRST\\
NGC 3656  &  4.5$\times$3.8 & 0.037 & 9.4 & 254 & LY07\\
NGC 4459  & 5.4$\times$5.4 & 0.15 & 0.86 & 5.7 & FIRST\\
NGC 4476  & 4.5$\times$4.5 & 0.17 & $<$ 0.5 & $<$ 3 & LY07\\
NGC 4526  & 5.4$\times$5.4 & 0.15 & 3.75 & 25 & FIRST\\
NGC 5666  & 4.6$\times$4.4 & 0.050 & 1.9 & 38 & LY07\\
\enddata
\tablecomments{
Columns 4 and 5 give the peak surface brightness and peak signal-to-noise ratio
on the target galaxy.
References: \citet{lucero} (LY07) or the VLA FIRST survey \citep{FIRST}.
}
\end{deluxetable}

\begin{deluxetable}{lrrrr}
\tablewidth{0pt}
\tablecaption{MIPS Flux Densities and Color Temperatures
\label{fluxes}}
\tablehead{
\colhead{Galaxy} & \colhead{$S_{24 \micron}$} & \colhead{$S_{70 \micron}$}
&
\colhead{$S_{160 \micron}$} & \colhead{Color Temp.} \\
\colhead{} & \colhead{Jy} & \colhead{Jy} & \colhead{Jy}  & \colhead{K}
}
\startdata
UGC 1503 & 0.0482 \error\ 0.0019 & 0.793 \error\ 0.079 & 1.49 \error\ 0.18 & 25.0 \error\ 0.7 \\
NGC 0807 & 0.0689 \error\ 0.0028 & 0.608 \error\ 0.061 & 2.39 \error\ 0.29 & 22.2 \error\ 0.5 \\
NGC 2320 & 0.0220 \error\ 0.0009 & 0.385 \error\ 0.077 & 1.12 \error\ 0.13 & 22.9 \error\ 0.5 \\
NGC 3032 & 0.165 \error\ 0.007 & 2.66 \error\ 0.266 & 3.32  \error\ 0.40 & 27.6 \error\ 0.6 \\
NGC 3656 & 0.169 \error\ 0.007 & 3.63 \error\ 0.36 & 3.81 \error\ 0.46 & 27.5 \error\ 0.8 \\
NGC 4459 & 0.129 \error\ 0.005 & 3.10 \error\ 0.19 & 3.76 \error\ 0.45 & 27.1 \error\ 0.6 \\
NGC 4476 & 0.0400 \error\ 0.0016 & 0.886 \error\ 0.133 & 1.38 \error\ 0.17 & 25.6 \error\ 0.7 \\
NGC 4526 & 0.314 \error\ 0.013 & 10.2 \error\ 0.61 & 12.7 \error\ 1.5 & 26.5 \error\ 0.5 \\
NGC 5666 & 0.162 \error\ 0.006 & 3.00 \error\ 0.30 & 2.48 \error\ 0.30 & 28.0 \error\ 0.8 \\
\enddata
\end{deluxetable}

\begin{deluxetable}{lccccc}
\tablewidth{0pt}
\tablecaption{FIR and Radio Continuum Emission
\label{q}}
\tablehead{
\colhead{Galaxy} & \colhead{$S_{350\micron}$} & \colhead{$S_{100 \micron}$} & \colhead{$S_{60 \micron}$}
&
\colhead{$S_{1.4 GHz}$} & \colhead{$q$} \\
\colhead{} & \colhead{Jy} & \colhead{Jy} & \colhead{Jy} & \colhead{mJy} & \colhead{}
}
\startdata
UGC 1503 & 0.24 \error\ 0.05 & 1.44 \error\ 0.15 & 0.50 \error\ 0.04 & 2.5 \error\ 0.5 & 2.57
\error\ 0.09 \\
NGC 0807 & 0.45 \error\ 0.09 & 1.83 \error\ 0.12 & 0.41 \error\ 0.03 & 1.2 \error\ 0.4 & 2.91
\error\ 0.15 \\
NGC 2320 & \nodata & 1.60 \error\ 0.16 & 0.26 \error\ 0.02 & 19.3 \error\ 0.7 & 1.60
\error\ 0.03 \\
NGC 3032 & \nodata & 4.70 \error\ 0.47 & 1.94 \error\ 0.10 & 7.2 \error\ 0.5 & 2.66
\error\ 0.04 \\
NGC 3656 & 0.69 \error 0.14 & 6.58 \error\ 0.66 & 2.54 \error\ 0.13 & 19.8 \error\ 0.7 & 2.35
\error\ 0.03 \\
NGC 4459 & \nodata & 4.82 \error\ 0.48 & 1.87 \error\ 0.09 & 1.8 \error\ 0.2  & 3.25
\error\ 0.05 \\
NGC 4476 & 0.28 \error 0.06 & 1.84 \error\ 0.18 & 0.66 \error\ 0.05 & $<$ 0.5 & $>$ 3.38  \\
NGC 4526 & \nodata & 17.1 \error\ 1.7 & 5.56 \error\ 0.05 & 12.0 \error\ 0.5 & 2.94
\error\ 0.03 \\
NGC 5666 & 0.52 \error 0.10 & 3.98 \error\ 0.40 & 1.99 \error\ 0.10 & 17.5 \error\ 0.7 & 2.24
\error\ 0.03 \\
\enddata
\tablecomments{Flux densities at 350\micron\ are taken from \citet{leeuw08}.
{\it IRAS} 60\micron\ and 100\micron\ flux densities are taken from the 1994
communication of Knapp as tabulated in NED.  The 20cm radio continuum fluxes 
are from the NRAO VLA Sky Survey \citep[NVSS;][]{NVSS} for the five targets
brightest at 20cm, and for NGC~4459 it is taken 
from the FIRST survey \citep{FIRST}.
The remaining three have 20cm radio continuum flux densities from \citet{lucero}.  
}
\end{deluxetable}

% schematic table here

\begin{deluxetable}{lcccccl}
\tablewidth{0pt}
\tabletypesize{\footnotesize}
\tablecaption{Summary of Star Formation Evidence\label{schematic}}
\tablehead{
\colhead{} & \multicolumn{3}{c}{Spatial Correlations} & \colhead{} & 
\colhead{} & \colhead{} \\
\colhead{Name} & \colhead{Resolved} & \colhead{24\micron\ vs.} & 
\colhead{24\micron\ } & \colhead{Optical/UV} & \colhead{FIR/radio} & 
\colhead{Notes}\\
\colhead{}   & \colhead{radio?} & \colhead{radio?} & \colhead{vs.\ CO?} & 
\colhead{} & \colhead{} & \colhead{}
}
\startdata
UGC~1503 & Y & Y & Y & \nodata & Y & SF ring. \\
NGC~0807 & Y & plateau/disk: Y & Y  & \nodata & Y & Some SF. \\
         &   & pt src: No      & No &         &   &          \\
NGC~2320 & No & No & No & \nodata & No & No SF?\\
NGC~3032 & Y & Y & Y & 1, 2, 3 & Y & SF inner disk.\\
NGC~3656 & Y & Y & Y & \nodata & Y & SF inner disk.\\
NGC~4459 & No & 2\asec\ disk: Y & Y & 1,2 & Y & SF inner disk.\\
         &    & 33\asec\ disk: No & No &  &   &               \\
NGC~4476 & \nodata & \nodata & Y & 4, 5 & \nodata & SF ring. \\
NGC~4526 & Y & plateau: Y & Y & 1, 2 & Y & SF inner disk.\\
         &   & 28\asec\ disk: No & No & &  &               \\
NGC~5666 & Y & Y & Y & 6 & Y & SF ring.\\
\enddata
\tablecomments{Column 2 refers to whether the cm-wave radio continuum emission is resolved
on size scales (and orientations) similar to those of the molecular disk.  Column
3 refers to whether the 24\micron\ emission or its morphological sub-components
have size scales similar to those of the radio continuum.  Column 4 refers to
whether the 24\micron\ emission morphology matches that of the molecular gas
(possibly being somewhat more compact than the molecular gas, if only part of the
gas is forming stars).
In column 5, references for the optical and UV star formation evidence are (1) \citet{sarzi06}; (2)
\citet{kuntschner06}; (3) \citet{mcdermid06a}; (4) this paper (Section \ref{4476}); (5) \citet{gdp07};
(6) \citet{dd03}.  Column 6 lists whether the FIR/radio flux density ratios are
consistent with star formation (Section~\ref{radioFIR}).}
\end{deluxetable}

\begin{deluxetable}{lc}
\tablewidth{0pt}
\tablecaption{24\micron\ Fit Parameters\label{fittable}}
\tablehead{
\colhead{Parameter} & \colhead{Value}
}
\startdata
\cutinhead{UGC~1503 (Ring)}
Total flux density & 46.6 \error 1.3 mJy \\
Radius of maximum & 4.8\asec \error 0.4\asec\ (1.65 \error 0.14 kpc) \\
Outer scale length & 3.2\asec \error 0.2\asec\ (1.10 \error 0.07 kpc) \\
Inner scale length & 2.2\asec \error 0.6\asec\ (0.8 \error 0.2 kpc) \\
Axis ratio & 0.73 \error 0.03 \\
Position Angle & 58\deg \error 2\deg \\

\cutinhead{NGC~0807 (Plateau, disk, and point source)}
Plateau outer scale  & 3.77\asec \error 0.12\asec\ (1.21 \error 0.04 kpc) \\
Plateau axis ratio & 0.423 \error 0.002 \\
Plateau PA & $-37.57\deg$ \error 0.06\deg \\
\\
Exp. disk flux density & 31.23 \error 0.07 mJy \\
Exp. disk scale length & 16.99\asec \error 0.14\asec\ (5.44 \error 0.04 kpc) \\
Exp. disk axis ratio & 0.577 \error 0.004 \\
Exp. disk PA & $-40.2\deg$ \error 0.2\deg \\
\\
Nuclear point source & 3.36 \error 0.08 mJy \\

\cutinhead{NGC~2320 ($r^{1/4}$ model)}
Total flux density & 20 \error 2 mJy \\
Effective radius (major) & 4.06\asec \error 0.06\asec\  (1.65 \error 0.02 kpc) \\
Effective radius (minor) & 1.88\asec \error 0.06\asec\ (0.77 \error 0.2 kpc) \\
PA & $-37\deg$ \error 2\deg \\

\tablebreak
\cutinhead{NGC~3032 (Exp. disk plus point source)}
Exp. disk flux density & 140 \error 10 mJy \\
Exp. disk scale length & 1.83\asec \error 0.04\asec\ (0.188 \error 0.004 kpc) \\
Exp. disk axis ratio & 0.808 \error 0.006 \\
Exp. disk PA & 63\deg \error 6\deg \\
\\
Off-nuclear point source\tablenotemark{a} & 6 \error 14 mJy \\

\cutinhead{NGC~3656 (Exponential disk)}
Exp. disk flux density & 146 \error 4 mJy \\
Exp. disk scale length & 2.0\asec \error 0.4\asec\ (0.39 \error 0.08 kpc) \\
Exp. disk axis ratio & 0.30 \error 0.05 \\
Exp. disk PA & $-5.7\deg$ \error 1.4\deg \\

\cutinhead{NGC~4459 (Two exponentials)}
Disk 1 flux density & 100 \error 1 mJy \\
Disk 1 scale length & 2.08\asec \error 0.02\asec\ (0.162 \error 0.001 kpc) \\
Disk 1 axis ratio & 0.706 \error 0.006 \\
Disk 1 PA & $-75.3\deg$ \error 0.7\deg \\
\\
Disk 2 flux density & 29 \error 5 mJy \\
Disk 2 scale length & 33\asec \error 8\asec\ (2.6 \error 0.6 kpc) \\
Disk 2 axis ratio & 0.85 \error 0.03 \\
Disk 2 PA & $-44\deg$ \error 2\deg \\

\cutinhead{NGC~4476 (Ring plus point source)}
Ring total flux density & 31.6 \error 0.2 mJy \\
Radius of maximum & 6.0\asec \error 0.2\asec\ (0.51 \error 0.02 kpc) \\
Outer scale length & 1.29\asec \error 0.07\asec\ (0.110 \error 0.006 kpc) \\
Inner scale length & 3.7\asec \error 0.6\asec\ (0.32 \error 0.05 kpc) \\
Axis ratio & 0.447 \error 0.009 \\
PA & 24.5\deg \error 0.6\deg \\
\\
Off-nuclear point source\tablenotemark{a} & 5 \error 2 mJy\\

\cutinhead{NGC~4526 (Plateau plus exponential disk)}
Plateau total flux density & 254 \error 1 mJy \\
Plateau radius & 7.5\asec \error 0.1\asec\ (0.629 \error 0.003 kpc) \\
Outer scale length & 1.91 \error 0.06\asec\ (0.160 \error 0.005 kpc) \\
Axis ratio & 0.20099 \error 0.00017 \\
PA & $-68.46\deg$ \error 0.14\deg \\
\\
Exp. disk flux density & 54.0 \error 0.6 mJy \\
Scale length & 28.4\asec \error 1.2\asec\ (2.4 \error 0.1 kpc) \\
Axis ratio & 0.397 \error 0.004 \\
PA & $-71\deg$ \error 0.4\deg \\

\cutinhead{NGC~5666 (Plateau)}
Total flux density & 144\error 3 mJy \\
Plateau radius & 5.87\asec \error 0.12\asec\ (0.88 \error 0.02 kpc) \\
Outer scale length & 0.5\asec \error 0.4\asec\ (0.08 \error 0.06 kpc) \\
Axis ratio & 0.86 \error 0.26 \\
PA & $-49\deg$ \error 28\deg \\
\enddata
\tablenotetext{a}{As the point source in NGC 3032 is weak and is 8\asec\ from
the galaxy center it may be a background source.  Similarly, the point source
in NGC 4476 is 6\asec \ northeast of the nucleus and has the effect of making
that ansa brighter, but it may also be a background source.}
\end{deluxetable}

\begin{deluxetable}{lccccr}
\tablewidth{0pt}
\tablecaption{Star Formation Rates and Depletion Timescales
\label{SFRs}}
\tablehead{
\colhead{Galaxy} & \colhead{log $L(24 \micron)$} & \colhead{SFR} &
\colhead{$\tau_{gas}$} \\
\colhead{} & \colhead{erg s$^{-1}$} & \colhead{\solmass~yr$^{-1}$} & \colhead{$10^9$
yr} 
}
\startdata
UGC 1503 & 42.57 & 0.60 & 4.3 \\
NGC 0807 & 42.66 & 0.69 & 2.7 \\
NGC 2320  & 42.37 & $<$ 0.40 & $>$ 17 \\
NGC 3032  & 42.05 & 0.21 & 3.2 \\
NGC 3656  & 42.61 & 0.66 & 7.7 \\
NGC 4459  & 41.71 & 0.083 & 2.8 \\
NGC 4476  & 41.27 & 0.043 & 3.2 \\
NGC 4526  & 42.15 & 0.22 & 4.0 \\
NGC 5666 & 42.37  & 0.40 & 1.5 \\
\enddata
\tablecomments{
% need to put this info into the text
%Column 2 gives the monochromatic luminosity at 24\micron, $L(24
%\micron) = 4\pi\; D^2\; S_{24} \nu_{24}$ where $D$ is the source distance (Table
%\ref{sampletable}), $S_{24}$ is the flux density at 24\micron\ and $\nu_{24}$ is
%the central frequency of the 24\micron\ band.  Column 3 gives the estimated star
%formation rate and column 4 the gas depletion timescale (see text).  
For NGC~2320
the star formation rate is an upper limit because we have argued that the bulk of
the 24\micron\ emission does not arise in star formation in this galaxy.  For
NGC~4459 and NGC~4526 we have argued that only $\approx$ 80\% of the 24\micron\
emission should be attributed to star formation and the star formation rate
reflects that attribution.  For NGC~807 we have argued that somewhere between 50\%
and 95\% of the 24\micron\ emission is driven by star formation activity but the
quoted star formation rate uses the higher value.
}
\end{deluxetable}

\clearpage %% too many unprocessed floats

\end{document}